\documentclass[lettersize,journal]{IEEEtran}
\usepackage{amsmath,amsfonts}
\usepackage{array}
\usepackage[caption=false,font=normalsize,labelfont=sf,textfont=sf]{subfig}
\usepackage{textcomp}
\usepackage{stfloats}
\usepackage{url}
\usepackage{verbatim}
\usepackage{graphicx}
\usepackage{cite}
\usepackage{graphicx}
\usepackage{amsmath}

\usepackage{amssymb}
\usepackage{booktabs}
\usepackage{url}

\usepackage{hyperref}
\usepackage{epsfig}
\usepackage{enumitem}
\setlist[itemize]{leftmargin=3mm}
\usepackage{multirow}
\usepackage{makecell}
\usepackage{xr}
\usepackage{enumitem}
\usepackage{tikz}
\usepackage[framemethod=tikz]{mdframed}

\usepackage{thmtools,thm-restate}

\usepackage[capitalize]{cleveref}
\crefname{section}{Sec.}{Secs.}
\Crefname{section}{Section}{Sections}
\Crefname{table}{Table}{Tables}
\crefname{table}{Tab.}{Tabs.}

\usepackage{tikz}
\usepackage{bbm}
\usepackage{dsfont}
\usepackage{mathrsfs}
\usepackage{amssymb}
\usepackage{amsmath}
\usepackage{amsthm}
\usepackage[ruled,vlined,linesnumbered]{algorithm2e}
\usepackage{algorithm2e}
\usepackage{algorithmicx} 
\usepackage{threeparttable}

\newcommand{\pie}[1]{%
\begin{tikzpicture}
 \draw (0ex,0ex) circle (1ex);
 \fill (0ex,-1ex) arc (-90:(#1-90):1ex) -- (0ex,-1ex) -- cycle;
\end{tikzpicture}%
}

\def\eg{\emph{e.g.,}\xspace}
\def\etc{\emph{etc}\xspace}
\def\ie{\emph{i.e.,}\xspace}
\def\etal{\emph{et al.}\xspace}

\begin{document}

\title{Bones of Contention: Exploring Query-Efficient Attacks against Skeleton Recognition Systems}

\author{Yuxin Cao\textsuperscript{*},
Kai Ye\textsuperscript{*},
Derui Wang, 
Minhui Xue, 
Hao Ge, 
Chenxiong Qian\textsuperscript{\dag},
Jin Song Dong\textsuperscript{\dag}
\thanks{Yuxin Cao and Jin Song Dong are with School of Computing, National University of Singapore, Singapore. Kai Ye and Chenxiong Qian are with School of Computing and Data Science, the University of Hong Kong, Hong Kong, China. Derui Wang and Minhui Xue are with CSIRO's Data61, Australia. Hao Ge is with Ping An Technology, China. 

\textsuperscript{*} These authors contributed equally.

\textsuperscript{\dag} Chenxiong Qian and 
Jin Song Dong are the corresponding authors. Emails:  cqian@cs.hku.hk, dcsdjs@nus.edu.sg.
}
}

\maketitle

\begin{abstract}
Skeleton action recognition models have secured more attention than video-based ones in various applications due to privacy preservation and lower storage requirements. Skeleton data are typically transmitted to cloud servers for action recognition, with results returned to clients via Apps/APIs. However, the vulnerability of skeletal models against adversarial perturbations gradually reveals the unreliability of these systems. Existing black-box attacks all operate in a decision-based manner, resulting in numerous queries that hinder efficiency and feasibility in real-world applications. Moreover, all attacks off the shelf focus on only restricted perturbations, while ignoring model weaknesses when encountered with non-semantic perturbations. In this paper, we propose two query-eff\underline{I}cient \underline{S}keletal \underline{A}dversarial \underline{A}tta\underline{C}ks, ISAAC-K and ISAAC-N. As a black-box attack, ISAAC-K utilizes Grad-CAM in a surrogate model to extract key joints where minor sparse perturbations are then added to fool the classifier. To guarantee natural adversarial motions, we introduce constraints of both bone length and temporal consistency. ISAAC-K finds stronger adversarial examples on the $\ell_\infty$ norm, which can encompass those on other norms. Exhaustive experiments substantiate that ISAAC-K can uplift the attack efficiency of the perturbations under 10 skeletal models. Additionally, as a byproduct, ISAAC-N fools the classifier by replacing skeletons unrelated to the action. We surprisingly find that skeletal models are vulnerable to large perturbations where the part-wise non-semantic joints are just replaced, leading to a query-free no-box attack without any prior knowledge. Based on that, four adaptive defenses are eventually proposed to improve the robustness of skeleton recognition models.
\end{abstract}


\section{Introduction}
As an important branch of computer vision, action recognition has been widely used in applications encompassing aged care~\cite{jang2020etri}, human-computer interaction~\cite{ann2014human}, video surveillance~\cite{han2018going}, intelligent transportation~\cite{xu2021action}, to name a few. In Metaverse systems, most virtual reality devices are equipped with joint sensors to obtain human posture in human-computer interaction~\cite{jiang2022avatarposer,dai2023sparse}. Compared to traditional RGB video action recognition methods~\cite{donahue2015long-term,wang2016TSN,hara2018R3D}, skeleton-based action recognition methods can greatly reduce the amount of computation and storage space, avoid the impact of illumination and occlusion~\cite{yue2022action,zheng2024towards}, and, more importantly, prevent privacy disclosure to some extent~\cite{moon2023anonymization}. Skeleton data can be easily obtained using joint sensors~\cite{jiang2022avatarposer,dai2023sparse} or devices like Kinect cameras~\cite{zhang2012microsoft}, positioning it as a new competitive modality. However, the robustness of skeletal models is not fully unveiled. Unlike pixel-level manipulations in images, perturbations in joint positions can preserve natural appearance yet still lead to severe semantic misclassification. Once compromised, such as failing to detect a fall incident in aged care, or misinterpreting suspicious behavior in a security context, these models may pose a grave threat to mission-critical environments and cause property loss and casualties. Hence, research into the robustness of skeleton recognition systems assumes paramount significance for ensuring secure and reliable applications.

\begin{figure*}[t]
\begin{center}
  \includegraphics[width=0.9\linewidth]{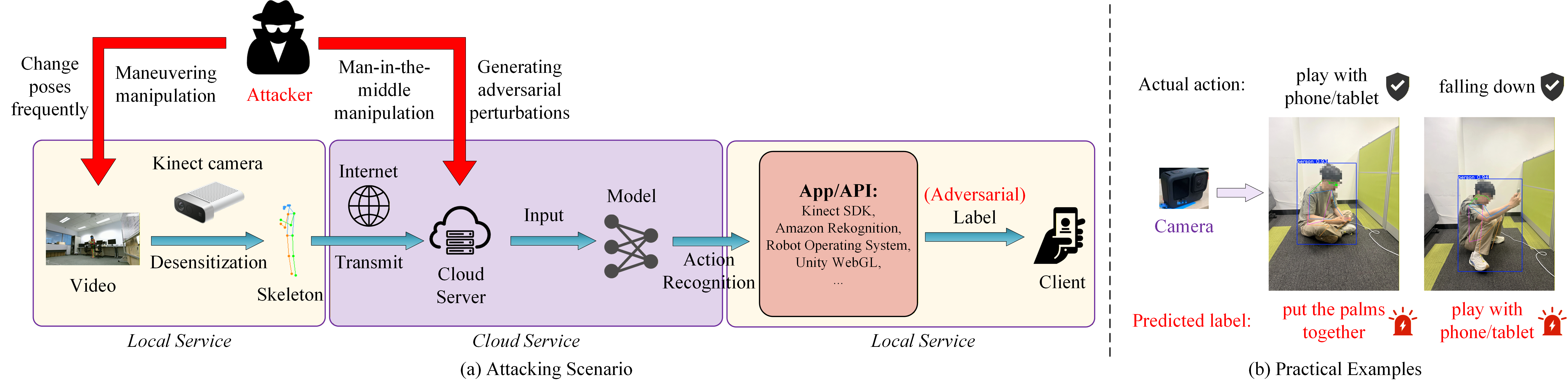}
\end{center}
\vspace{-3mm}
  \caption{(a) Attacking Scenario. A typical skeleton recognition system includes a local service (extracting skeleton from video and receiving system output through Apps/APIs) and a cloud service (storing the skeleton on a cloud server and inputting it into the recognition model). The attack can occur on both the cloud server and original videos. (b) Practical examples of misclassification.}
\label{fig:scenario}
\end{figure*}

The security scenario of this paper is illustrated in Fig.~\ref{fig:scenario}(a). Most recognition systems delegate data storage and model training/inference to outsourcing companies due to unbearable burden. Concretely, after the skeleton data are obtained by sensors, they are uploaded to the cloud server for action recognition. Finally, the output is sent to the client through Apps/APIs, \eg Kinect SDK and Amazon Rekognition. However, the reliance on third-party outsourcing services raises significant concerns about data privacy and security. We consider two ways of attacks in this process. 1) \textbf{Man-in-the-middle manipulation}: the attacker can intercept the skeleton received on the cloud server and replace it with elaborately designed adversarial examples. A strong motivation of this attack is to conceal urgent events captured by surveillance cameras, such as critical situations in aged care or illicit activities, by hindering these incidents from being easily detected or flagged. 
2) \textbf{Maneuvering manipulation}: the person whose pose is tracked (not necessarily the client) can act as an attacker.
For example, the perpetrators are aware that their behavior is monitored, but they can make an attempt to fool the recognition model to some extent by frequently changing non-semantic poses since we find that by simply replacing lower body with various postures, existing skeletal models tend to output different labels for the same upper-body action such as drinking water (see more details in Section~\ref{sec:experiment_ISAAC-N}).

To concretely demonstrate the feasibility and potential harm of such attacks, Fig.~\ref{fig:scenario}(b) shows practical cases where adversarial perturbations cause critical misclassifications in workplace monitoring and elderly care. For instance, actions such as sitting on the floor or falling are incorrectly recognized as benign gestures such as ``put the palms together'' or ``playing with phone/tablet'', potentially leading to undetected misconduct or missed emergency alerts. These examples reveal how adversarial skeleton manipulation, even under constrained visual cues, can undermine the trustworthiness of action recognition systems deployed in safety-critical environments.

\begin{table}[t]  
\centering
\caption{Comparison of adversarial attacks against skeleton action recognition models.}
\label{tab:study_comparison}
\vspace{-1mm}
\resizebox{0.99\linewidth}{!}{
\begin{threeparttable}
\begin{tabular}{c|cccc|cc|cc}
\toprule
\textbf{Approaches} & \textbf{White-box} & \textbf{Gray-box} & \textbf{Black-box} & \textbf{No-box} & \textbf{Dense} & \textbf{Sparse} & \textbf{Restricted} & \textbf{Unrestricted} \\
\midrule
Kumar~\cite{kumar2020finding} & \pie{360} & \pie{0} & \pie{0} & \pie{0} & \pie{0} & \pie{360} & \pie{360} & \pie{0} \\
Liu~\cite{liu2020adversarial} & \pie{360} & \pie{0} & \pie{0} & \pie{0} & \pie{360} & \pie{0} & \pie{360} & \pie{0} \\
Zheng~\cite{zheng2020towards}  & \pie{360} & \pie{0} & \pie{0} & \pie{0} & \pie{360} & \pie{0} & \pie{360} & \pie{0} \\
Wang~\cite{wang2021understanding}  & \pie{360} & \pie{360} & \pie{0} & \pie{0} & \pie{360} & \pie{0} & \pie{360} & \pie{0} \\
Tanaka~\cite{tanaka2022adversarial}  & \pie{360} & \pie{0} & \pie{0} & \pie{0} & \pie{360} & \pie{0} & \pie{360} & \pie{0} \\
Huang~\cite{huang2022sparse}  & \pie{360} & \pie{0} & \pie{0} & \pie{0} & \pie{0} & \pie{360} & \pie{360} & \pie{0} \\
BASAR~\cite{diao2021basar}  & \pie{0} & \pie{0} & \pie{360} & \pie{0} & \pie{360} & \pie{0} & \pie{360} & \pie{0}\\
QESAR~\cite{kang2023qesar}  & \pie{0} & \pie{0} & \pie{360} & \pie{0} & \pie{360} & \pie{0} & \pie{360} & \pie{0} \\
Lu~\cite{lu2023hard}  & \pie{0} & \pie{0} & \pie{0} & \pie{360} & \pie{360} & \pie{0} & \pie{360} & \pie{0} \\
TASAR~\cite{diao2024tasar} & \pie{0} & \pie{360} & \pie{0} & \pie{0} & \pie{360} & \pie{0} & \pie{360} & \pie{0} \\
\midrule
\makecell[c]{ISAAC-K (ours)} & \pie{0} & \pie{0} & \pie{360} & \pie{0} & \pie{360}\tnote{1} & \pie{360} & \pie{360} & \pie{0} \\
\makecell[c]{ISAAC-N (ours)} & \pie{0} & \pie{0} & \pie{360} & \pie{360} & \pie{360} & \pie{360} & \pie{360} & \pie{360} \\
\bottomrule
\end{tabular}
\begin{tablenotes}
\item[1] The variant of our attack supports dense perturbations. See more details in Section~\ref{sec:ISAAC-N}.
\end{tablenotes}
\end{threeparttable}
}
\vspace{-1mm}
\end{table}

The above scenarios drive us to search for adversarial perturbations for skeleton data.
Despite a line of existing skeleton action recognition attacks (as listed in Table~\ref{tab:study_comparison}), they are encountered with three main limitations. 
1) Most attacks are under a white-box setting~\cite{kumar2020finding,liu2020adversarial,zheng2020towards,wang2021understanding,tanaka2022adversarial,huang2022sparse}, which requires a wealth of prior knowledge of the victim model. 
However, in actual scenarios, most commercial models do not disclose their model architecture, so accessibility to the model is impractical, which necessitates the emergence of black-box attacks where attackers can only obtain the model output. Despite few gray-box attacks that rest upon the transferability of skeletal models, their performance is quite weak and susceptible to the selected surrogate model~\cite{diao2024tasar,sun2024invisibility}.
2) Existing black-box attacks are lacking in attack efficiency. BASAR~\cite{diao2021basar} and QESAR~\cite{kang2023qesar} are both decision-based black-box attacks, which require a large number of queries due to the search process.
Moreover, they perturb all joints, while the perturbations on some inconsequential joints are not necessary, which significantly reduces the attack efficiency. In practice, attack efficiency counts, since large queries imply high attack costs.
A later study~\cite{huang2022sparse} considers adding sparse perturbations to skeletons by extracting key joints with dynamic attention, but the attack is carried out under the white-box setting, which greatly overestimates the ability of the attacker.
Note that squarely transferring white-box methods to black-box attacks poses challenges. This is primarily due to two reasons. First, unlike images, the dimensions of skeleton data are much lower; thus, diminutive perturbations at any joint may cause optimization to fall into local optima. Additionally, the lack of gradient information can easily lead to an impasse in the search for perturbation directions, ultimately leading to the inability to converge.
3) All off-the-shelf attacks consider restricted attacks that only add minuscule perturbations to the input motion. 
We regard this as reasonable, but it is not the only way to add perturbations since the non-semantic joints have no direct connection with the classification label. 
For a sufficiently robust classifier, the non-semantic joints added with large perturbations are not expected to affect the classification results as long as the whole skeleton conforms to the human manifold. 
Taking an action classified as ``drinking water'' as an example, no matter how the lower body changes (\eg sitting, crouching, or kneeling), an expected classifier should always classify it as ``drinking water'', even if the skeleton sample has not appeared in the training set. 

To close the gaps above, we first propose 
a query-eff\underline{I}cient \underline{S}keletal \underline{A}dversarial \underline{A}tta\underline{C}k based on \underline{K}ey joints, called ISAAC-K. We only add perturbations to key joints, while the naturalness of adversarial motions is maintained by two domain-specific constraints. The spatial bone length constraint maintains kinematic plausibility by recursively enforcing bone length consistency across child joints and parent joints, while the temporal consistency constraint ensures perturbation smoothness across adjacent frames by dynamically correcting perturbation directions based on temporal alignment. 
Afterwards, the skeletons are optimized by an improved version of RayS~\cite{chen2020rays}, which substantially reduces the effective search complexity and enables faster optimization. Experiments show that ISAAC-K performs better than the existing open-source State-Of-The-Art (SOTA) attack, BASAR, and can bypass existing adversarial defenses. For non-semantic joints, following the semantic invariant perturbations considered in the image/video domain~\cite{shamsabadi2020colorfool,cao2023stylefool}, we propose ISAAC-N and show that, even in a no-box setting (the attacker cannot even query the model), by making reasonable changes to the non-semantic joints, the model output can be misguided without affecting the human perception of skeleton actions. This part-wise substitution approach deviates from conventional perturbation-based frameworks and exposes a new form of vulnerability: the model's over-reliance on part of the body. These perturbations, which surpass metrical restrictions, are marked as unrestricted perturbations~\cite{bhattad2020unrestricted}, significantly enhancing their attacking performance and resistance to adaptive defenses. The proposed ISAAC-K and ISAAC-N are customized for man-in-the-middle manipulation and maneuvering manipulation delineated above, respectively. They also help expose structural vulnerabilities and provide valuable insights to enhance model robustness for model developers.

Our contributions are summarized as follows:
\begin{itemize}
    \item We propose a novel query-efficient black-box skeletal adversarial attack, ISAAC-K. 
    We use Grad-CAM~\cite{selvaraju2017gradcam} to extract key joints from a local surrogate model and search for stronger adversarial examples on the $\ell_\infty$ ball, as it is the superset of other $\ell_p$ balls under the same perturbation radius.
    \item We tailor an efficient optimization process for generating adversarial perturbations by improving RayS, and consider both bone length (intra) and temporal consistency (inter) constraints, which helps maintain the naturalness of the adversarial motions.
    \item Experiments on three datasets and 10 skeletal models, including two state-of-the-art (SOTA) ViT-based models, show that
    ISAAC-K outperforms the existing open-source SOTA attack, BASAR, in terms of average query by a large margin while maintaining indistinguishability in terms of consistency, naturalness, and realness. 
    \item As a byproduct, ISAAC-N reveals that even when the attacker cannot query skeletal models, they are surprisingly vulnerable to semantic invariant perturbations such as \textit{part-wise} replacement of non-semantic joints. As responses, we propose four adaptive defenses to improve the robustness of skeletal models.
\end{itemize}

\section{Related Work}

\noindent\textbf{Skeleton Attacks.} Despite a plethora of action recognition methods based on skeletons~\cite{shi2020msaagcn,yan2018stgcn,xu2023lagcn,do2024skateformer}, 
research on adversarial attacks against such models is still in its infancy.
Most attacks against skeleton action recognition models are under the white-box setting, where the attacker can access the structure and parameters of the victim model. 
An early study~\cite{kumar2020finding} extracts key frames and key joints through feature information for multi-modal data including RGB videos, depth videos and skeleton data.
Recent skeleton attacks preserve the imperceptibility of the adversarial skeleton by restricting either the bone length~\cite{tanaka2022adversarial,liu2020adversarial}, rotation angle~\cite{zheng2020towards}, velocity~\cite{zheng2020towards}, or joint accelerations~\cite{liu2020adversarial,wang2021understanding}. However, such minor changes may cause off-manifold skeletons that violate the basic requirements of human dynamics. To handle this, a black-box attack called BASAR~\cite{diao2021basar} is proposed to move the perturbed skeleton to the manifold. 
This random search-based attack, however, causes a large number of queries and low attack efficiency, which are not mentioned in the paper. Kang \etal~\cite{kang2023qesar} claim that they reduce query to some extent by introducing bias sampling, but the accuracy of gradient estimation is affected by the sampling number, and the constraints are not clarified in detail.
Huang \etal~\cite{huang2022sparse} uses dynamic attention to select joints that are highly dynamic, but under a white-box setting. 
Moreover, the joints extracted by attention mechanism~\cite{kumar2020finding,huang2022sparse} are not consistent with key joints closely related to classification, and the latter plays a greater role in affecting model outputs.
Lu \etal~\cite{lu2023hard} propose a hard no-box attack in which the attacker cannot obtain the victim model and labels. However, the attack success rate remains low, which makes this difficult attack setting not effective. 
A recent work~\cite{diao2024tasar} explores the transferability of skeletal adversarial examples in the gray-box setting, but suffers from low performance and sensitivity to surrogate models.

\noindent\textbf{Skeleton Defenses.} There is also a lack of research on skeleton defenses, leaving space for adversaries to attack surreptitiously without being detected. Zheng \etal~\cite{zheng2020towards} design an effective plug-in defense by adding Gaussian noise and temporal filtering. Wang \etal~\cite{wang2023defending} propose a black-box defense, BEAT, which discriminates adversarial motions from the view of joint Bayesian together with a perturbation restriction strategy based on natural motion manifolds. BEAT can successfully resist BASAR and outstrip competitors such as Zheng \etal~\cite{zheng2020towards} and adversarial training (AT)~\cite{madry2018towards} by improving the classifier's robustness and reducing the memory footprint. Hu \etal~\cite{hu2023self} devise a graph convolution network for multilevel skeleton forgery detection to recognize anomalies, but it is not applicable to adversarial perturbations. Recently, Tanaka \etal~\cite{tanaka2024fourier} tailor a Fourier-based AT method for $\ell_2$-PGD attack to improve model robustness.

\section{Preliminary}
\begin{figure}[t]
\begin{center}
  \includegraphics[width=0.98\linewidth]{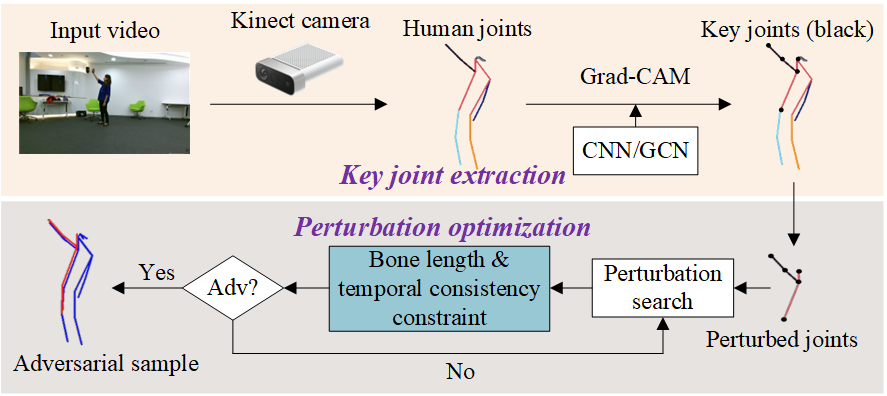}
\end{center}
\vspace{-3mm}
  \caption{Overview of our proposed ISAAC-K.}
\label{fig:framework}
\vspace{-1mm}
\end{figure}

\noindent\textbf{Problem Statement.}
We denote the skeleton action recognition model as $f\left( {\mathbf{x}} \right):\mathbf{x} \to y$, which takes a motion $\mathbf{x}\in {\mathbb{R}^{T \times N \times D}}$ as input, and outputs the predicted label $y$, where $T$, $N$, and $D$ represent the frame number, degree of freedoms (usually positions or angles of joints) and joint dimension, respectively. The attacker then regards the attack as an optimization problem by solving:
\begin{equation}\label{equ:goal}
\small
\begin{array}{c}
\mathop {\arg \min }\limits_{{\mathbf{x}_{adv}}} \hat{L}\left( {\mathbf{x},{\mathbf{x}_{adv}}} \right), \quad \\
\mathrm{s.t.} \quad {\mathbf{x}_{adv}} \in \left[ {0,1} \right], \ \ {f\left( {{\mathbf{x}_{adv}}} \right) \ne y},
\end{array}
\end{equation}
where $\hat{L}$ denotes the distance function which ensures that the adversarial motion $\mathbf{x}_{adv}$ is close to the original motion $\mathbf{x}$. Note that it is common practice to rescale the data to $\left[ {0,1} \right]$ before feeding it into models. The adversarial motion is expected to be predicted as any class excluding the ground truth class $y$ in untargeted attacks, while the pre-determined class in targeted attacks. Solving the above non-linear problem is difficult and requires substitutional means. The previous work, BASAR~\cite{diao2021basar}, utilizes a numerical solution and moves the adversarial motion to the human manifold $\mathcal{M}$ to maintain naturalness. In our work, we modify the above optimization function and adapt it to two different sparse attacks, ISAAC-K and ISAAC-N.
ISAAC-K denotes the sparse attack in which perturbations are only added to the key joints, the framework of which is depicted in Fig.~\ref{fig:framework}. 
The key joints are first extracted by a local surrogate model, after which the perturbations are optimized iteratively under the constraints of bone length and temporal consistency.
As a derivative of ISAAC-K, ISAAC-N represents the sparse attack in which perturbations are added to non-semantic joints, which will be introduced in Section~\ref{sec:ISAAC-N}.

\noindent\textbf{Threat Model.}
We assume that the attacker mounts the attack under black-box settings, where there is no access to the architecture and parameters of the victim model. More strictly, we consider hard-label attacks (consistent with BASAR~\cite{diao2021basar}), which indicates that the attacker can only obtain the Top-1 label when querying the model. We also set an upper query limit, which emulates the constraint imposed by commercial action recognition systems in the real world.
Besides, in ISAAC-N, the attacker can replace the non-semantic joints by uploading the specific action taken by himself or from other online resources, instead of selecting from the dataset. This enables action customization for users and supports specified skeleton attacks.

\section{Methodology}
In this section, we introduce ISAAC-K and ISAAC-N in detail. ISAAC-K can be boiled down to key joint extraction and perturbation optimization. The optimization strategy is customized to preserve the naturalness of adversarial skeletons. Moreover, ISAAC-N explores the possibility of query-free no-box attacks by replacing \textit{part-wise} non-semantic joints.

\subsection{Key Joint Extraction}
Key joints refer to those closely related to higher-level features that can directly affect the classification results. To reduce the perturbations added to the motion and improve efficiency, we need to extract the key joints for the victim model. Since the attacker has no access to the inner features of the model, we circuitously harness a local surrogate model to extract key joints and transfer them to the black-box victim model based on the fact that the deep features learned from different models have certain transferability~\cite{jiang2019black}.
Here we draw on Grad-CAM~\cite{selvaraju2017gradcam} and adapt it to skeleton data. 
The significance score for each class is derived from the gradients of the feature maps before the output layer. Specifically, the weight of the $k$-th feature map corresponding to class $c$ at the $l$-th layer is computed as: 
\begin{equation}\label{equ:weight}
\small
\alpha_k^{l,c}=\frac{1}{N}\sum_{i=1}^{N}\frac{\partial y^c}{\partial F_{k,i}^l},
\end{equation}
where $y^c$ denotes the confidence score of class $c$, and $F_{k,i}^l$ represents the feature at the $i$-th joint in the $k$-th feature map at the $l$-th layer. 
Then, the significance score is calculated as:
\begin{equation}\label{equ:heatmap}
\small
{L^c}[l,i] = \mathrm{ReLU}\left( {\sum\limits_k {\alpha _k^{l,c}F_{k,i}^l\left( {\mathbf{x},A} \right)} } \right),
\end{equation}
where $\mathrm{ReLU}$ represents the ReLU activation function, and $A$ is the adjacency matrix. 
Finally, joints are ranked according to their significance scores, and the top $N_k$ joints are selected as the \textit{key joints}. 
We hope that sparse perturbations are applied only to the same joints in each frame, since it would be more practical in real-world scenarios. Though Grad-CAM may output different key joints among different frames, we use the top $N_k$ joints sorted in descending order of count for all key joints extracted from $T$ frames.
We conduct a pilot experiment to verify the minor frame-wise key-joint difference. Given 100 videos from NTU60 RGB+D~\cite{shahroudy2016ntu60} and a pre-trained ST-GCN~\cite{yan2018stgcn}, we calculate the average cosine similarity of the one-hot version of key joints (\eg $\left[0,1,1,0,0,1,...,1 \right]^T$) from different frames. The result turns out to be 0.9719, indicating that it is convincing to use the count-based key joint selection strategy mentioned above. 

Overall, Grad-CAM helps us find several joints that are crucial to surrogate model prediction. Perturbing these joints is more likely to mislead black-box models due to model transferability. 
Note that the surrogate model serves only as a tool for extracting key joints rather than as a local white-box model used to generate transferable examples, as in traditional transfer-based attacks~\cite{papernot2016transferability}. Further analyses in Section~\ref{subsec:surrogate} show that our method is insensitive to the choice of surrogate model when selecting key joints. Even when attacking a proprietary commercial system, using an open-source skeleton model to extract key joints is completely applicable.

\subsection{Perturbation Optimization}\label{subsec:perturbation_optimization}
Existing hard-label black-box attacks typically perform gradient estimation in a continuous search space using finite difference or zeroth order optimization~\cite{cheng2019query,cheng2020signopt}. However, each gradient update requires multiple rounds of calculating the decision boundary radius which is estimated through binary search. This results in a prohibitively large number of queries, making continuous-space optimization inefficient and impractical under hard-label constraints. 
It is empirically found that adversarial examples generated by gradient-based attacks are found mostly in the vertices of the $\ell_\infty$ ball~\cite{moon2019parsimonious}, which is similar to the phenomenon in high-dimensional Gaussian that the majority of the probability mass is distributed around the spherical shell~\cite{bishop2006pattern}. Besides, with the same radius, the $\ell_\infty$ ball is a superset of other $\ell_p$ balls. Therefore, searching on the $\ell_\infty$ ball can find stronger adversarial examples~\cite{chen2020frank,moon2019parsimonious}. To the best of our knowledge, RayS~\cite{chen2020rays} has been proposed and verified as one of the most efficient hard-label attacks under black-box setting in the image domain, which reduces the search space from $\mathbb{R}^d$ to $\{ -1, 1\}^d$ by searching only from the ray direction space, where $d$ denotes the input dimension ($d = T \times N \times D$ for skeleton data).
With RayS, we propose the $\ell_\infty$-based black-box attack against skeleton action recognition models. 
Instead of Equation~\ref{equ:goal}, we turn to solving: 
\begin{equation}\label{equ:optmization1}
\small
\mathop {\min }\limits_{{\mathbf{d}} \in {{\left\{ { - 1, 1} \right\}}^d}} {g}\left( {\mathbf{d}} \right),
\end{equation}
where \small $g\left( {\mathbf{d}} \right) = \mathop {\arg \min }\limits_r \mathbbm{1}\left\{ {f\left(  {\mathbf{x} + r\frac{{\mathbf{d}}}{{{{\left\| {\mathbf{d}} \right\|}_2}}}} \right) \ne y} \right\}$ 
\normalsize denotes the decision boundary distance obtained by binary search, a classical strategy employed in decision-based attacks~\cite{cheng2019query,chen2020rays}. $r\in\mathbb{R}^ + $ is a multiplier that controls the perturbation size.

Despite the high efficiency of RayS, squarely transferring RayS from images to skeletons faces the problem of naturalness loss, since solely searching perturbations from the vertices of $\ell_\infty$ ball might lead to artifacts that are contrary to human kinematics. 
In order to maintain the naturalness of the adversarial motion, we introduce an intra constraint and an inter constraint.

\begin{figure}[t]
\begin{center}
  \includegraphics[width=0.7\linewidth]{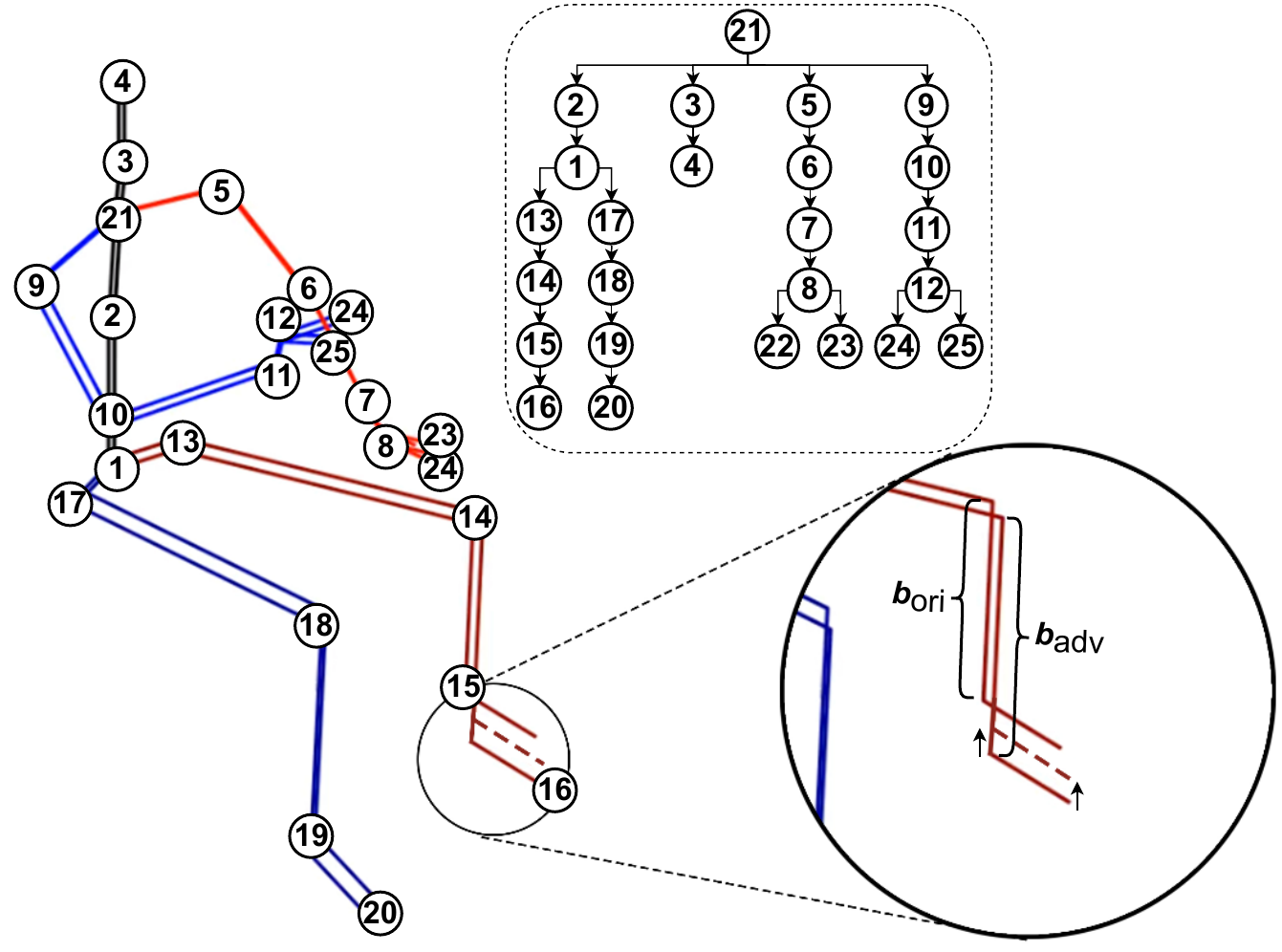}
\end{center}
\vspace{-3mm}
  \caption{Sketch map of bone length constraint.}
\label{fig:bone_constraint}
\vspace{-1mm}
\end{figure}

\noindent\textbf{Bone Length (Intra) Constraint.~}
It is low-efficient to restrict the joint position by the distance function in Equation~\ref{equ:goal} since additional velocity and acceleration restriction may also be necessitated to avoid artifacts. Instead, we restrict the bone length between the original skeleton and the adversarial skeleton to achieve a more effective and efficient attack. A prior work~\cite{tanaka2022adversarial} has shown that perturbing bone length is more realizable. 
To concretely demonstrate the relation between child joints and parent joints in the bone length constraint, Fig.~\ref{fig:bone_constraint} shows the 25 joints in a skeleton from the NTU60 RGB+D dataset~\cite{shahroudy2016ntu60}. We abstract all joints of the skeleton into a tree, with the root joint being the central joint of the skeleton (joint 21 in the NTU dataset). Then we search recursively from the root joint each time. If the change in a bone relative to its original length exceeds a clipping threshold, we modify its child joints according to the tree in Fig.~\ref{fig:bone_constraint}. 
For the $i$-th adversarial joint ${{\mathbf{x}}_i^{'}}$, its position is adjusted as follows.
\begin{equation}\label{equ:bone_cons}
\small
\begin{array}{l}
{\mathbf{x}}_i^{'} \leftarrow {\mathbf{x}}_p^{'} + \frac{{{{\mathbf{d}}_{adv}}}}{{{{\left\| {{{\mathbf{d}}_{adv}}} \right\|}_2}}}{\mathbf{\tilde d}},\\
{\mathbf{\tilde d}} = {\rm{Clip}}\left( {{{\left\| {{{\mathbf{d}}_{adv}}} \right\|}_2},\left( {1 - {\varepsilon _b}} \right){{\left\| {{{\mathbf{d}}_{ori}}} \right\|}_2},\left( {1 + {\varepsilon _b}} \right){{\left\| {{{\mathbf{d}}_{ori}}} \right\|}_2}} \right),
\end{array}
\end{equation}
where $\mathbf{x}_p^{'}$ denotes the parent joint corresponding to ${{\mathbf{x}}_i^{'}}$, ${{\mathbf{d}}_{ori}} = {{\mathbf{x}}_i} - {{\mathbf{x}}_p}$, ${{\mathbf{d}}_{adv}} = {\mathbf{x}}_i^{'} - {\mathbf{x}}_p^{'}$, $\rm{Clip}$ denotes the clip operation to limit the input, ${{\varepsilon _b}}$ represents the bone length constraint threshold.
In short, we restrict the bone length to a reasonable range by pulling its child joints based on the linkage mechanism between child joints and parent joints.

\begin{figure}[t]
\begin{center}
  \includegraphics[width=0.9\linewidth]{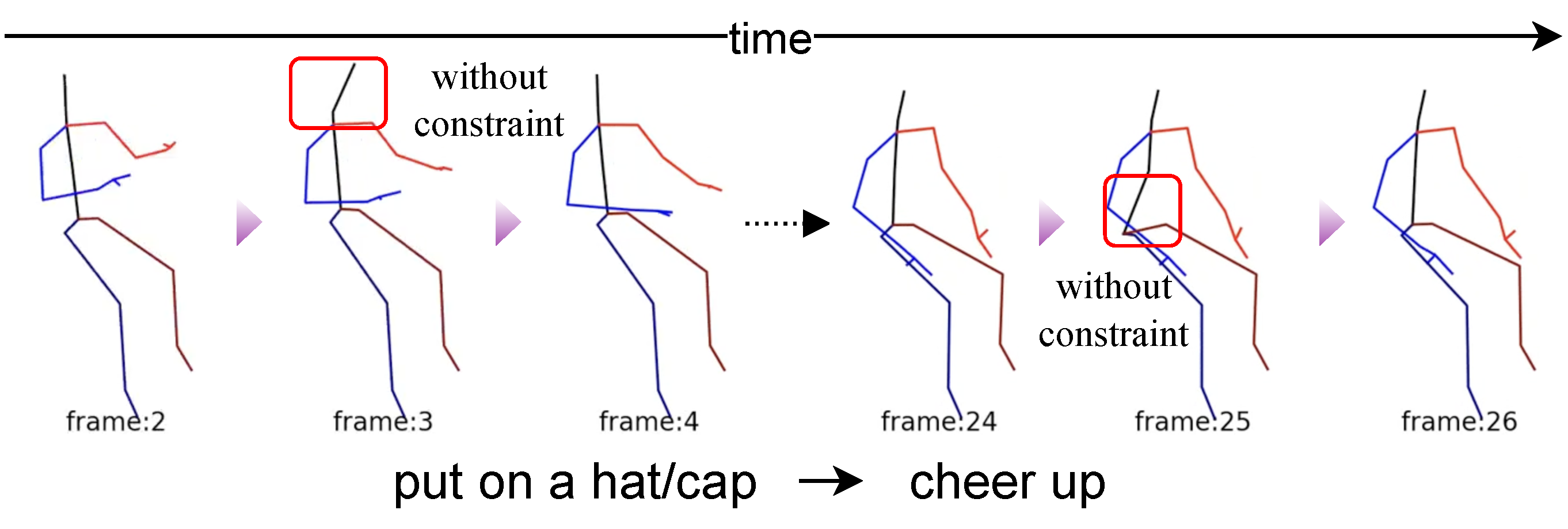}
\end{center}
\vspace{-5mm}
  \caption{An example of adversarial skeletons when temporal consistency constraint is not considered.}
\label{fig:temporal_consistency_constraint}
\vspace{-1mm}
\end{figure}

\noindent\textbf{Temporal Consistency (Inter) Constraint.}
We pose the temporal consistency constraint based on two reasons. 1) As skeleton data are expressed only by joint positions, minor perturbations may cause apparent flickers or anomalous variations between consecutive frames, as shown in Fig.~\ref{fig:temporal_consistency_constraint}. The abrupt changes between adjacent frames make adversarial skeletons easily detectable. 2) The hierarchical version of RayS~\cite{chen2020rays}, which can improve the attack efficiency over the naive version, first splits the search direction into $2^s$ ($s$ denotes the stage in~\cite{chen2020rays}) blocks and then changes the sign of each block simultaneously. It has no impact on images, but significantly affects the temporal consistency since the search directions of some joints in consecutive frames are totally opposite when (very frequently) the block consists of the last few joint positions in the first frame and the first few joint positions in the next frame. Therefore, not only can restricting the bone length of consecutive frames of the adversarial motion maintain temporal naturalness, but also be an improvement of RayS when applied to time series data. To be specific, we define the Temporal Directional Consistency (TDC) to measure the frame-to-frame consistency of perturbation directions of the skeleton. The TDC (range: $[0,1]$) between two adjacent frames 
$k-1$ and $k$ is computed as:
\begin{equation}
\text{TDC}^{[k]} = \frac{1}{2} * {cosine\_{similarity}}\left( \hat{\mathbf{d}}^{[k-1]},\ \hat{\mathbf{d}}^{[k]} \right) + \frac{1}{2},
\end{equation}
where $\hat{\mathbf{d}}^{[k]} = \mathbf{x}'^{[k]} - \mathbf{x}^{[k]}$ denotes the perturbation direction at frame $k$. 
Then we define a soft interpolation weight using an exponential decay function:
\begin{equation}
\lambda^{[k]} = \exp(-\alpha \cdot \text{TDC}^{[k]}),
\end{equation}
where $\alpha$ is a large integer that ensures that $\lambda^{[k]}$ is close to 0 when $\text{TDC}^{[k]}$ reaches 1. In other words, when the two directions are already similar (\textit{i.e.,} $\text{TDC}^{[k]} \to 1$), the weight $\lambda^{[k]}$ becomes negligible. This property allows the update to largely retain the current perturbation. In contrast, when the directions differ significantly (\textit{i.e.,} $\text{TDC}^{[k]} \to 0$), the weight $\lambda^{[k]}$ is close to 1, indicating that the correction prioritizes the previous frame to avoid abrupt temporal changes. 
Therefore, the adversarial skeleton is updated by interpolating between the current and previous perturbation directions:
\begin{equation}
\mathbf{x}'^{[k]} \leftarrow \mathbf{x}^{[k]} + \lambda^{[k]} \cdot \hat{\mathbf{d}}^{[k-1]} + (1 - \lambda^{[k]}) \cdot \hat{\mathbf{d}}^{[k]}.
\end{equation}
This update adaptively steers the current perturbation direction toward the previous one based on their alignment and reduces unnatural frame-to-frame variations in the adversarial motion.

\noindent\textbf{Objective Function.}
We define the decision boundary distance with the input motion restricted by bone length and temporal consistency as follows.
\begin{equation}
\small
g_{res}\left( {\mathbf{d}} \right) = \mathop {\arg \min }\limits_r \mathbbm{1}\left\{ {f\left( {\Theta \left( {\mathbf{x} + r\frac{{\mathbf{d}}}{{{{\left\| {\mathbf{d}} \right\|}_2}}},\mathbf{x},{\varepsilon _b}} \right)} \right) \ne y} \right\},
\end{equation}
where $\Theta $ denotes the perturbation constraint operation of bone length and temporal consistency. Then, we turn to solving
\begin{equation}
\small
\mathop {\min }\limits_{{\mathbf{d}} \in {{\left\{ { - 1, 1} \right\}}^{\rho d}}} {g_{res}}\left( {M \odot \mathbf{d}} \right),
\end{equation}
where $M$ denotes the flattened key joint mask matrix, $\odot$ stands for the Hadamard product, $\rho = N_k/N$ represents the joint sparsity.
The sparse perturbation can reduce the search space to $\{ -1, 1\}^{\rho d}$, where the adversarial motion is selected from the vertices of parts of the $\ell_\infty$ ball. 

\subsection{ISAAC-N}\label{sec:ISAAC-N}
As a byproduct, we find replacing \textit{part-wise} non-semantic joints can also bring threats to skeletal models. Therefore, we propose ISAAC-N. For non-key joints obtained in the key joint extraction, we believe that some of these joints are relatively insensitive to the model, \ie adding perturbations to the non-semantic parts has a relatively small impact on the recognition results. Even from the perspective of human visual perception, the perturbations of non-semantic parts will not affect human subjective judgment, \eg as for the action of ``picking up the mobile phone and taking selfies'', we may pay more attention to whether the arm and head are raised visually than the posture or action of the lower body. If perturbing the joints of the lower body to a large extent can cause misclassification of the model, we deem that the model is unsafe in this context. This is based on the basic fact that a robust model is expected to always focus on the arm joints regardless of changes in the lower body, \ie standing, sitting, prostrating.

Taking the NTU120 RGB+D dataset~\cite{liu2019ntu120} as an example, we find that most of the non-semantic regions are in the lower body, which is also consistent with human cognition since most of the classes in this dataset are upper-body actions. We simply design four typical lower body postures, sit, crouch, kneel, and half kneel, to replace the lower body of the input motion, followed by root joint translation and normalization. If the replaced motion is not adversarial yet, we follow the improved RayS-based perturbation optimization mentioned in Section~\ref{subsec:perturbation_optimization}.
Note that the temporal consistency constraint helps to improve temporal stability, which is particularly evident in ISAAC-N when large perturbations are considered. Therefore, we provide three attacking strategies for ISAAC-N: only replacing the lower body (ISAAC-NR), replacement followed by adding perturbations to only the lower body (ISAAC-NRL) and all joints (ISAAC-NRA). 

\section{Experimental Evaluation}
In this section, we begin with an experimental setup, followed by a comprehensive evaluation including quantitative and qualitative analyses of ISAAC-K and ISAAC-N, ablation study and countermeasures to show the overall performance of our proposed methods. Finally, we discuss the extension to targeted attacks and analyze the role of the surrogate model.

\subsection{Experimental Setup}
\noindent\textbf{Datasets.}
We use three prevalently used datasets, NTU60 RGB+D (NTU60)~\cite{shahroudy2016ntu60}, NTU120 RGB+D (NTU120)~\cite{liu2019ntu120} and NW-UCLA~\cite{wang2014nw_ucla}. NTU60 was captured by three Kinect V2 cameras in 2016 and comprised of 56,880 video clips with 60 classes performed by 40 subjects. The dataset contains 40 everyday actions, 11 two-person interactions, and 9 health actions. 
Extended by NTU60, NTU120 was released 3 years later and was captured by the same cameras, with another 57,600 video clips and 60 classes, among which 42 classes are everyday actions, 15 classes are two-person interactions, and 3 classes are health actions. The videos were performed by 106 different subjects. 
NW-UCLA includes 1,494 video clips with 10 action classes. These videos were captured using three Kinect V1 cameras (20 body joints) positioned at various viewpoints. Of these views, two are used for training and one for testing. 
As NTU60 and NTU120 both contain 2D (17 body joints) and 3D (25 body joints) data, we use both of them in our experiments. 
Other datasets such as MHAD~\cite{ofli2013mhad}, PKU-MMD~\cite{liu2017pkummd} are smaller Kinect-based datasets with similar characteristics. Therefore, we do not include them in our experiments. The HDM05 dataset~\cite{muller2007hdm05}, released in 2007, is outdated, limited in scale, and prone to motion-capture artifacts, making it unsuitable for modern evaluations. Therefore, our experiments already validate the method's performance under the standard and most representative data setting.

\noindent\textbf{Models.}
We carefully select victim models to show our attack's effectiveness and scalability.
For NTU60 and NTU120, we first select six GCN (Graph Convolutional Network)-based models (MS-AAGCN~\cite{shi2020msaagcn}, CTR-GCN~\cite{chen2021ctrgcn}, DG-STGCN~\cite{duan2022dgstgcn}, MS-G3D~\cite{liu2020msg3d}, ST-GCN~\cite{yan2018stgcn} and ST-GCN++~\cite{duan2022stgcn++}) due to their representativeness. We also involve two SOTA GCN-based models (TD-GCN~\cite{liu2023tdgcn} and LA-GCN~\cite{xu2023lagcn}) for NW-UCLA. 
In recent years, the emergence of Vision Transformer (ViT) has significantly enhanced classification performance in skeleton action recognition tasks. Therefore, we choose two SOTA accessible ViT-based 3D models (Hyperformer~\cite{zhou2022hypergraph} and SkateFormer~\cite{do2024skateformer}) for all datasets. 
We train these models from scratch unless the pre-trained models are provided publicly from a public toolbox, PYSKL~\cite{pyskl}.

\noindent\textbf{Competitors.}
We compare our method with BASAR~\cite{diao2021basar} since it is the only existing open-source attack under the black-box setting. BASAR uses a random search to find the adversarial motion in the human manifold. 
For a fair comparison, we modify BASAR from $\ell_2$ to $\ell_\infty$. We will provide the justification when introducing the metrics later. 
We do not compare our attack with the later proposed black-box attack, QESAR~\cite{kang2023qesar}, for the following reasons. 1) QESAR is not open-source, and its reproduction is challenging due to insufficiently clarified constraints in its paper. 2) QESAR employs a decision-based strategy similar to BASAR, which is time-consuming due to its reliance on random search. As a result, although QESAR claims to reduce the query number compared to BASAR, it still requires an order of 1,000 queries to achieve success, whereas our attack only requires an order of 10 queries (see Section~\ref{subsec:attack_performance}).

\noindent\textbf{Attack Preliminaries.}
For each datasets, we randomly select 500 skeletal motions. We use ST-GCN as our local surrogate model to extract key joints for NTU 60/120. The key joint number is set as 12/18 for 2D/3D data due to better overall performance. For NW-UCLA, we adopt SkateFormer as the surrogate model and select 14 key joints accordingly.
We set the restriction threshold $\varepsilon$ (the $\ell_\infty$ norm between original and adversarial motion) as 0.4 by comprehensively considering the stealthiness-efficiency trade-off. 
See Section~\ref{subsec:ablation} for more details about different key joint numbers and different $\varepsilon$. Following~\cite{tanaka2022adversarial}, we set the bone length constraint threshold $\varepsilon _b$ as 0.1. 
The decay constant $\alpha$ is set as 10 and the query limit is set as 2,000.

\noindent\textbf{Metrics.}
We use Attack Success Rate (ASR) and Average Query (AQ) to evaluate the attack performance. 
ASR denotes the ratio of adversarial motions that successfully mislead the classifier. Any attack exceeding the query limit (2,000) will also be considered a failure. AQ stands for the required average query number for successful attacks. 
When ASR is close, lower AQ becomes more significant. 
To gauge the distance of joints towards root joint, we also introduce the averaged off-centered deviation $l_c$, which can be expressed as:  
\begin{equation}
\small
{l_c} = \frac{1}{n}\sum\limits_{i = 1}^{n} {\sum\limits_{j = 1,j \ne c}^N {\left| {{{\left\| {{{\mathbf{x}}_{\left( {i,j} \right)}} - {{\mathbf{x}}_{\left( {i,c} \right)}}} \right\|}_2} - {{\left\| {{\mathbf{x}}_{\left( {i,j} \right)}^{'} - {\mathbf{x}}_{\left( {i,c} \right)}^{'}} \right\|}_2}} \right|} },
\end{equation}
where ${{{\mathbf{x}}_{\left( {i,j} \right)}}}$ denotes the $j$-th joint of the $i$-th motion,
$c$ is the root joint of the skeleton we select (\eg joint 21 in the NTU dataset) 
and $n$ is the number of adversarial motions. 
As a complement to the metrics, following previous works~\cite{zheng2020towards,diao2021basar}, we use the averaged joint acceleration deviation $\Delta a$ to quantitatively measure the difference between original and adversarial motions: 
\begin{equation}
\small
\Delta a=\frac{1}{nTN} \sum_{i=1}^{n} \left \| \ddot{\mathbf{x}}_{(i)}-\ddot{\mathbf{x}}^{'}_{(i)}  \right \|_2 ,
\end{equation}
where $\ddot{\mathbf{x}}_{(i)}$ denotes the joint acceleration deviation of the $i$-th motion. 
$l_c$ and $\Delta a$ describe the static and dynamic changes of bones, respectively. 
We replace the averaged joint position deviation $l$ used in BASAR with $l_c$ for two reasons. On the one hand, the skeletons may be translated as a whole in the image, resulting in an inaccurate calculation of the absolute distance in $l$. On the other hand, it is unfair to compare the $\ell_2$ distance in $l$ since we consider $\ell_\infty$ restriction in our attack. Therefore, we calculate the average distance between each joint to the root joint to ensure a fair comparison as much as possible.

\begin{table}[t]
    \centering   
    \caption{Attack performance comparison on NTU60.}
    \label{tab:attack_performance_ntu60}
    \vspace{-1mm}
    \resizebox{0.98\linewidth}{!}{
    \begin{tabular}{ccrrrrrrrrrrrrrrrrrrrrrrrrr}
    \toprule
    \multirow{2}{*}[-0.5ex]{\makecell[c]{\textbf{Model}\\(\textbf{Test Accuracy})}} & 
    \multirow{2}{*}[-0.5ex]{\textbf{Attack}} & \multicolumn{4}{c}{\textbf{NTU60(2D)}} & \multicolumn{4}{c}{\textbf{NTU60(3D)}} \\
    \cmidrule(r){3-6}\cmidrule(r){7-10}
    & & \footnotesize{\textbf{ASR}$\uparrow$} & \footnotesize{\textbf{AQ}$\downarrow$} & \footnotesize{\textbf{$l_c$}} & \footnotesize{\textbf{$\Delta a$}} 
    & \footnotesize{\textbf{ASR}$\uparrow$} & \footnotesize{\textbf{AQ}$\downarrow$} & \footnotesize{\textbf{$l_c$}} & \footnotesize{\textbf{$\Delta a$}} \\
    \midrule
    \multirow{2}{*}{\makecell[c]{MS-AAGCN\\(89.7\%/89.0\%)}} & BASAR & 0.99 & 51.48 & 0.08 & 2.92 
    & 0.97 & 133.58 & 0.10 & 3.06  \\
     & ISAAC-K & \textbf{1.00} & \textbf{22.32} & 0.13 & 2.34 
     & \textbf{0.99} & \textbf{29.55} & 0.05 & 2.41    \\ 
    \midrule
    \multirow{2}{*}{\makecell[c]{CTR-GCN\\(90.6\%/89.6\%)}} & BASAR & \textbf{1.00} & 84.32 & 0.08 & 3.65 
    & 0.92 & 152.22 & 0.10 & 3.52 \\
     & ISAAC-K  & \textbf{1.00} & \textbf{20.55} & 0.13 & 2.46 
     & \textbf{0.99} & \textbf{27.44} & 0.05 & 2.01  \\
    \midrule
    \multirow{2}{*}{\makecell[c]{DG-STGCN\\(N/A/91.2\%)}} & BASAR & N/A
    & N/A & N/A & N/A 
    & 0.94 & 160.34 & 0.08 & 4.36 \\
     & ISAAC-K  & N/A & N/A & N/A & N/A 
     & \textbf{0.99} & \textbf{38.31} & 0.08 & 3.02 \\
    \midrule
    \multirow{2}{*}{\makecell[c]{MS-G3D\\(92.7\%/89.6\%)}} & BASAR & 0.98 & 67.04 & 0.09 & 2.86 
    & 0.89 & 91.93 & 0.08 & 2.90 \\
     & ISAAC-K  & \textbf{1.00} & \textbf{23.86} & 0.09 & 2.07 
     & \textbf{0.99} & \textbf{27.25} & 0.09 & 2.10  \\
    \midrule
    \multirow{2}{*}{\makecell[c]{ST-GCN\\(89.0\%/87.8\%)}} & BASAR & 0.98 & 39.17 & 0.08 & 2.85 
    & 0.93 & 134.74 & 0.10 & 3.14  \\
     & ISAAC-K  & \textbf{1.00} & \textbf{17.43} & 0.04 & 0.92 
     & \textbf{1.00} & \textbf{20.26} & 0.03 & 0.88 \\
    \midrule
    \multirow{2}{*}{\makecell[c]{ST-GCN++\\(89.3\%/89.3\%)}} & BASAR& 0.97 & 50.02 & 0.09 & 3.42 
    & 0.92 & 172.14 & 0.10 & 2.87  \\
     & ISAAC-K  & \textbf{1.00} & \textbf{19.23} & 0.10 & 2.06 
     & \textbf{1.00} & \textbf{34.71} & 0.11 & 2.39  \\
    \midrule
    \multirow{2}{*}{\makecell[c]{Hyperformer\\(N/A/90.8\%)}} & BASAR & N/A & N/A & N/A & N/A & 0.64 & 49.34 & 0.07 & 2.38\\
    & ISAAC-K & N/A & N/A & N/A & N/A &\textbf{0.89} &\textbf{28.31} & 0.05 & 1.32 \\
    \midrule
    \multirow{2}{*}{\makecell[c]{SkateFormer\\(N/A/92.6\%)}} & BASAR & N/A & N/A & N/A & N/A & 0.60 & 93.85 & 0.07 & 3.39 \\
    & ISAAC-K & N/A & N/A & N/A & N/A  &\textbf{0.88} &\textbf{75.49} & 0.24 & 6.25 \\
    \bottomrule
    \end{tabular}
    }
    \vspace{-1mm}
\end{table}
\begin{table}[t]
    \centering   
    \caption{Attack performance comparison on NTU120.}
    \label{tab:attack_performance_ntu120}
    \vspace{-1mm}
    \resizebox{0.98\linewidth}{!}{
    \begin{tabular}{ccrrrrrrrrrrrrrrrrrrrrrrrrr}
    \toprule
    \multirow{2}{*}[-0.5ex]{\makecell[c]{\textbf{Model}\\(\textbf{Test Accuracy})}} & 
    \multirow{2}{*}[-0.5ex]{\textbf{Attack}} & \multicolumn{4}{c}{\textbf{NTU120(2D)}}  & \multicolumn{4}{c}{\textbf{NTU120(3D)}}\\
    \cmidrule(r){3-6}\cmidrule(r){7-10}
    & & \footnotesize{\textbf{ASR}$\uparrow$} & \footnotesize{\textbf{AQ}$\downarrow$} & \footnotesize{\textbf{$l_c$}} & \footnotesize{\textbf{$\Delta a$}}  
    & \footnotesize{\textbf{ASR}$\uparrow$} & \footnotesize{\textbf{AQ}$\downarrow$} & \footnotesize{\textbf{$l_c$}} & \footnotesize{\textbf{$\Delta a$}} \\
    \midrule
    \multirow{2}{*}{\makecell[c]{MS-AAGCN\\(80.2\%/82.8\%)}} & BASAR 
    & 0.79 & 24.77 & 0.09 & 2.72 
    & 0.79 & 108.57 & 0.10 & 2.85   \\
     & ISAAC-K
     & \textbf{1.00} & \textbf{18.75} & 0.06 & 1.64 
     & \textbf{0.97} & \textbf{30.73} & 0.08 & 2.23    \\ 
    \midrule
    \multirow{2}{*}{\makecell[c]{CTR-GCN\\(82.2\%/84.0\%)}} & BASAR 
    & 0.60 & 37.11 & 0.08 & 2.67 
    & 0.93 & 107.48 & 0.11 & 2.40 \\
     & ISAAC-K  
     & \textbf{0.99} & \textbf{20.05} & 0.08 & 2.17 
     & \textbf{0.98} & \textbf{34.11} & 0.09 & 2.33 \\
    \midrule
    \multirow{2}{*}{\makecell[c]{DG-STGCN\\(N/A/85.7\%)}} & BASAR 
    & N/A & N/A & N/A & N/A 
    & 0.93 & 215.48 & 0.10 & 3.47 \\
     & ISAAC-K 
     & N/A & N/A & N/A & N/A 
     & \textbf{1.00} & \textbf{43.42} & 0.09 & 2.16 \\
    \midrule
    \multirow{2}{*}{\makecell[c]{MS-G3D\\(85.5\%/84.0\%)}} & BASAR 
    & \textbf{1.00} & 36.32 & 0.07 & 3.26 
    & 0.75 & 123.98 & 0.10 & 3.05 \\
     & ISAAC-K  
     & \textbf{1.00} & \textbf{24.27} & 0.04 & 1.64 
     & \textbf{0.98} & \textbf{35.29} & 0.07 & 2.21 \\
    \midrule
    \multirow{2}{*}{\makecell[c]{ST-GCN\\(80.1\%/82.1\%)}} & BASAR 
    & 0.66 & 34.23 & 0.09 & 3.14 
    & 0.98 & 117.02 & 0.09 & 2.98 \\
     & ISAAC-K  
     & \textbf{1.00} & \textbf{16.89} & 0.04 & 1.07 
     & \textbf{1.00} & \textbf{22.60} & 0.05 & 1.40 \\
    \midrule
    \multirow{2}{*}{\makecell[c]{ST-GCN++\\(84.4\%/83.2\%)}} & BASAR
    & \textbf{1.00} & 34.41 & 0.11 & 3.19 
    & 0.92 & 131.18 & 0.10 & 2.64 \\
     & ISAAC-K  
     & \textbf{1.00} & \textbf{20.47} & 0.06 & 1.77 
     & \textbf{0.98} & \textbf{30.26} & 0.05 & 1.15 \\
    \midrule
    \multirow{2}{*}{\makecell[c]{Hyperformer\\(N/A/70.2\%)}} & BASAR & N/A & N/A & N/A & N/A & 0.80 & 176.71 & 0.07 & 1.51 \\
    & ISAAC-K & N/A & N/A & N/A & N/A &\textbf{0.88} &\textbf{21.71}  & 0.10 & 1.18 \\
    \midrule
    \multirow{2}{*}{\makecell[c]{SkateFormer\\(N/A/86.6\%)}} & BASAR & N/A & N/A & N/A & N/A & 0.75 & 114.17 & 0.10 & 5.89\\
    & ISAAC-K & N/A & N/A & N/A & N/A &\textbf{0.83} &\textbf{47.63} & 0.17 & 2.86 \\
    \bottomrule
    \end{tabular}
    }
    \vspace{-1mm}
\end{table}
\begin{table}[t]
    \centering   
    \caption{Attack performance comparison on NW-UCLA.}
    \label{tab:attack_performance_nwucla}
    \vspace{-1mm}
    \resizebox{0.6\linewidth}{!}{
    \begin{tabular}{ccrrrrrrrrrrrrrrrrrrrrrrrrr}
    \toprule
    \multirow{2}{*}[-0.5ex]{\makecell[c]{\textbf{Model}\\(\textbf{Test Accuracy})}} & 
    \multirow{2}{*}[-0.5ex]{\textbf{Attack}} & \multicolumn{4}{c}{\textbf{NW-UCLA}} \\
    \cmidrule(r){3-6}
    & & \footnotesize{\textbf{ASR}$\uparrow$} & \footnotesize{\textbf{AQ}$\downarrow$} & \footnotesize{\textbf{$l_c$}} & \footnotesize{\textbf{$\Delta a$}} \\
    \midrule
    \multirow{2}{*}{\makecell[c]{TD-GCN\\(93.8\%)}} & BASAR & 0.58 & 193.13 & 0.05 & 7.42 \\
     & ISAAC-K & \textbf{0.91} & \textbf{22.76} & 0.07 & 6.66  \\ 
    \midrule
    \multirow{2}{*}{\makecell[c]{LA-GCN\\(95.7\%)}} & BASAR &0.64  & 320.21 & 0.03 & 12.03 \\
     & ISAAC-K & \textbf{1.00} & \textbf{31.64} & 0.07 & 7.83  \\ 
    \midrule
    \multirow{2}{*}{\makecell[c]{Hyperformer\\(93.8\%)}} & BASAR & 0.71 & 205.05  & 0.04 & 8.32 \\
     & ISAAC-K &  \textbf{0.91} &  \textbf{32.05} & 0.03 & 3.60  \\ 
    \midrule
    \multirow{2}{*}{\makecell[c]{SkateFormer\\(94.8\%)}} & BASAR & 0.14 & 328.5  & 0.07 & 10.38\\
     & ISAAC-K & \textbf{0.97} & \textbf{49.71} & 0.08 & 4.18  \\ 
    \bottomrule
    \end{tabular}
    }
    \vspace{-1mm}
\end{table}

\subsection{Attack Performance of ISAAC-K}\label{subsec:attack_performance}
\begin{figure*}[t]
\begin{center}
  \includegraphics[width=0.9\linewidth]{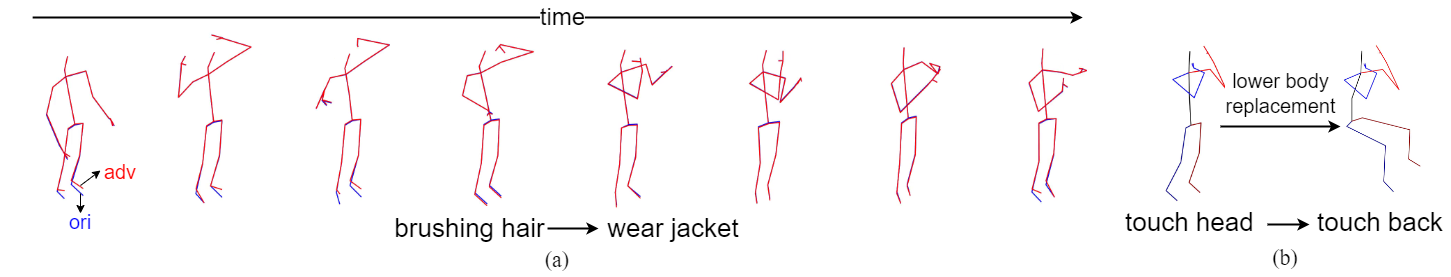}
\end{center}
\vspace{-3mm}
\caption{(a) An example of ISAAC-K with minor perturbations ($\ell_\infty=0.17$) added to key joints. (b) An example of ISAAC-NR. 
}
\label{fig:vis}
\vspace{-2mm}
\end{figure*}

\begin{figure}[t]
\begin{center}
  \includegraphics[width=1\linewidth]{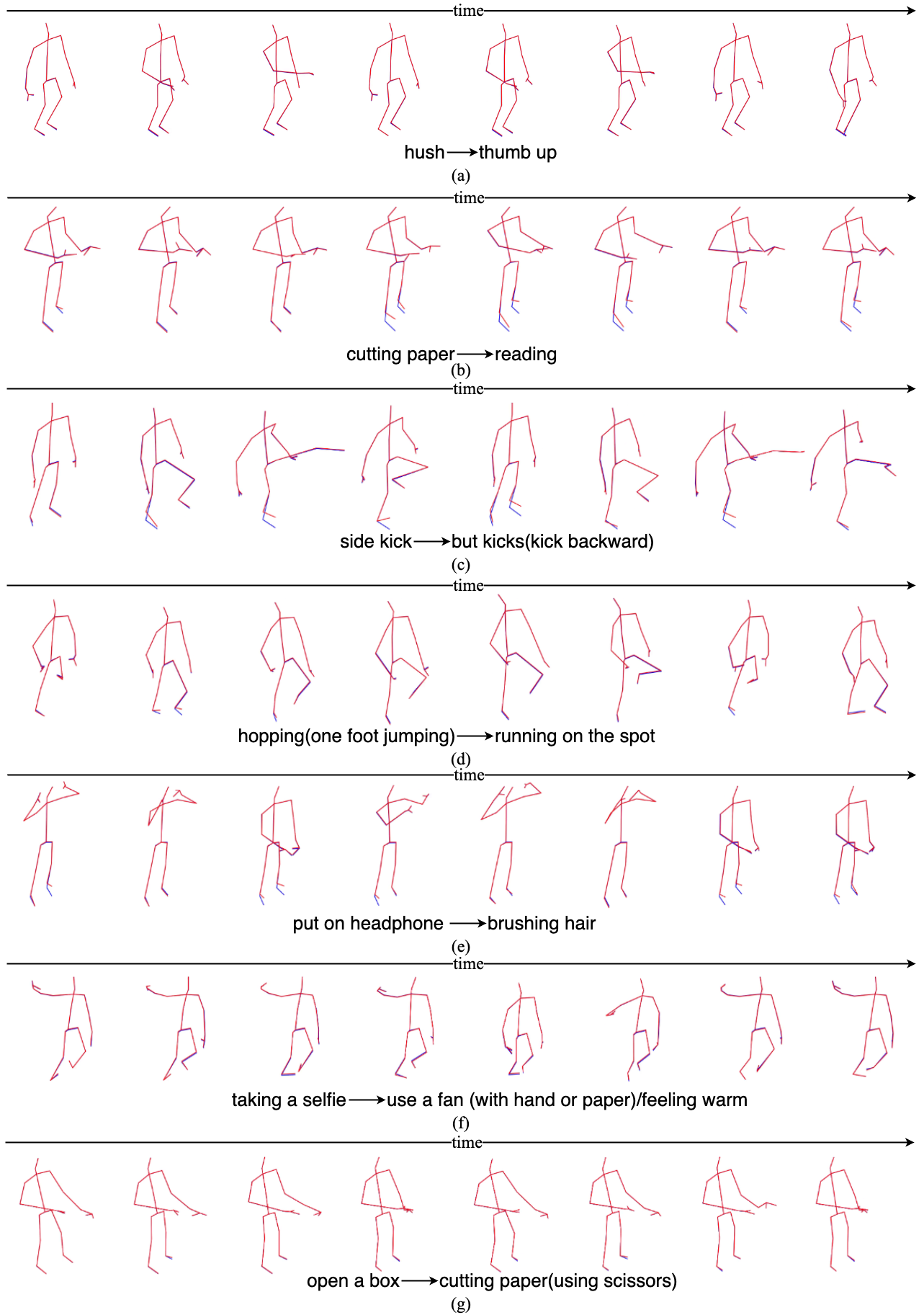}
\end{center}
\vspace{-3mm}
\caption{Some examples of ISAAC-K with minor perturbations ($\ell_\infty<0.4$) added to key joints. Blue and red skeletons denote original motions and adversarial motions, respectively.}
\label{fig:example_isaac_k}
\end{figure}

\noindent\textbf{Attack Results.}
Table~\ref{tab:attack_performance_ntu60},~\ref{tab:attack_performance_ntu120} and~\ref{tab:attack_performance_nwucla} report the attack performance of BASAR and ISAAC-K on NTU60, NTU120 and NW-UCLA respectively. 
The accuracy for the test data is shown under the model name, suggesting that the models are well-trained. 
In most cases, the ASR of ISAAC-K achieves nearly 100\%. However, ISAAC-K consumes much fewer queries than BASAR, indicating that ISAAC-K is more query-efficient. This superiority can be attributed to a smaller search space than boundary-based attack and joint sparsity, where only key joints that have a strong correlation with the output label are perturbed. Although BASAR considers the on-manifold issue, the solution of human body dynamics often falls into a stalemate, resulting in the inability to converge. 
In terms of quantitative metrics regarding stealthiness, ISAAC-K also performs better than BASAR on $l_c$ and $\Delta a$ in most cases. 
Due to the sparsity of perturbations, a successful attack may require the addition of slightly larger perturbations to a limited number of joints, resulting in some cases where the metrics are larger than those of BASAR. However, these figures are still small enough that the perturbations are difficult for human eyes to distinguish, which will be verified at length in Section~\ref{subsec:user_study}. Considering that ASR and AQ are the most important indicators to measure attack efficiency, we do not bold these two metrics in tables.
Fig.~\ref{fig:vis}(a) shows an example of the original and adversarial motions generated by ISAAC-K. The adversarial perturbation is rather small, and the stealthiness can be well-preserved. To verify the universality of this conclusion, Fig.~\ref{fig:example_isaac_k} provides more visualizations.

Moreover, although the key joints of ISAAC-K are extracted from ST-GCN, the competitive attack performance against unseen models indicates that the extracted key joints have a strong transferability across models.
Another phenomenon is that attacks against 2D skeletons are found to be easier (fewer queries) than 3D skeletons because of fewer body joints.

\begin{figure}[t]
\begin{center}
  \includegraphics[width=0.9\linewidth]{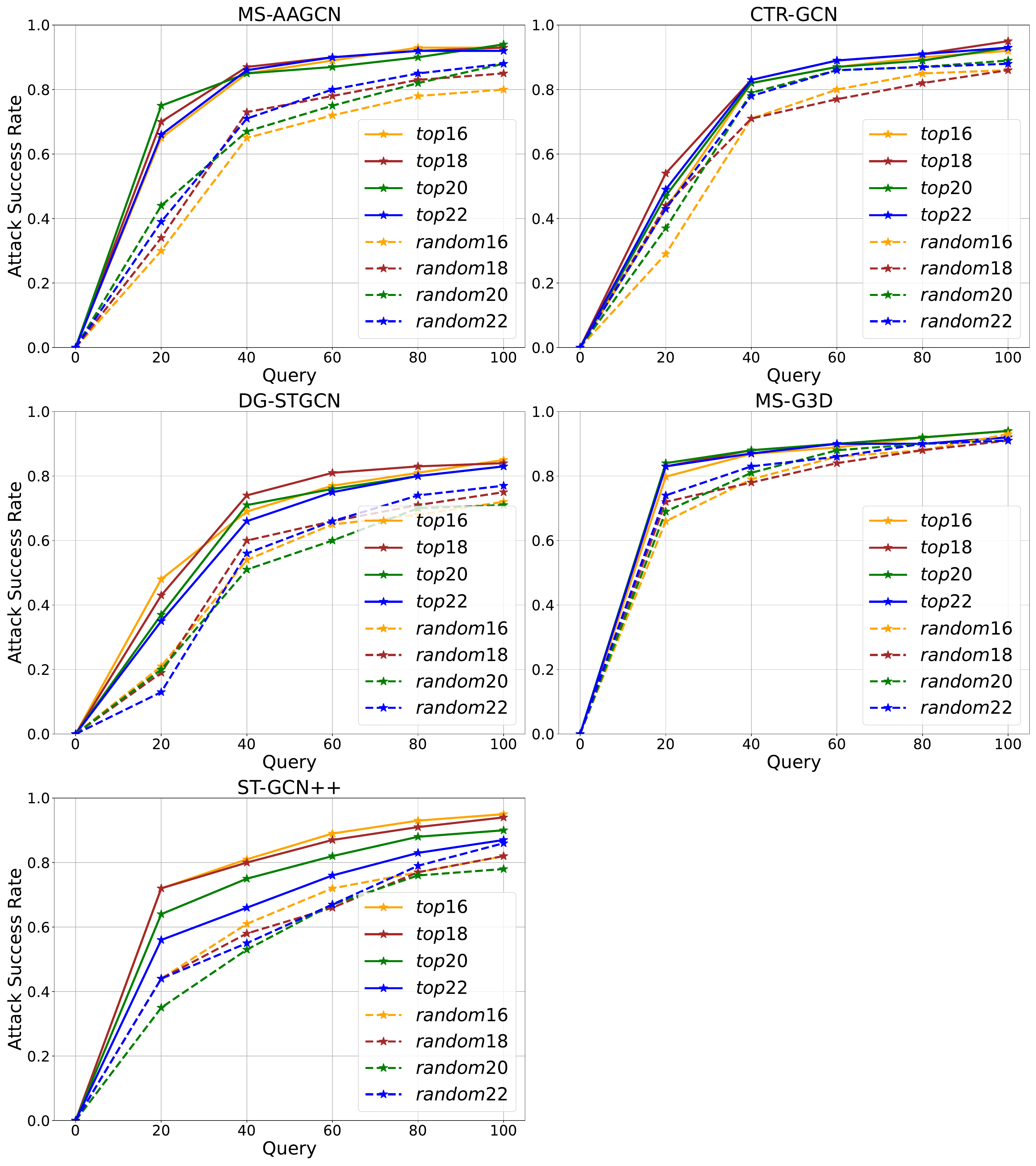}
\end{center}
\vspace{-3mm}
\caption{Attack performance under key joint selection and random joint selection across different joint numbers.}
\label{fig:key_joint}
\vspace{-2mm}
\end{figure}

\begin{figure}[t]
\begin{center}
  \includegraphics[width=0.9\linewidth]{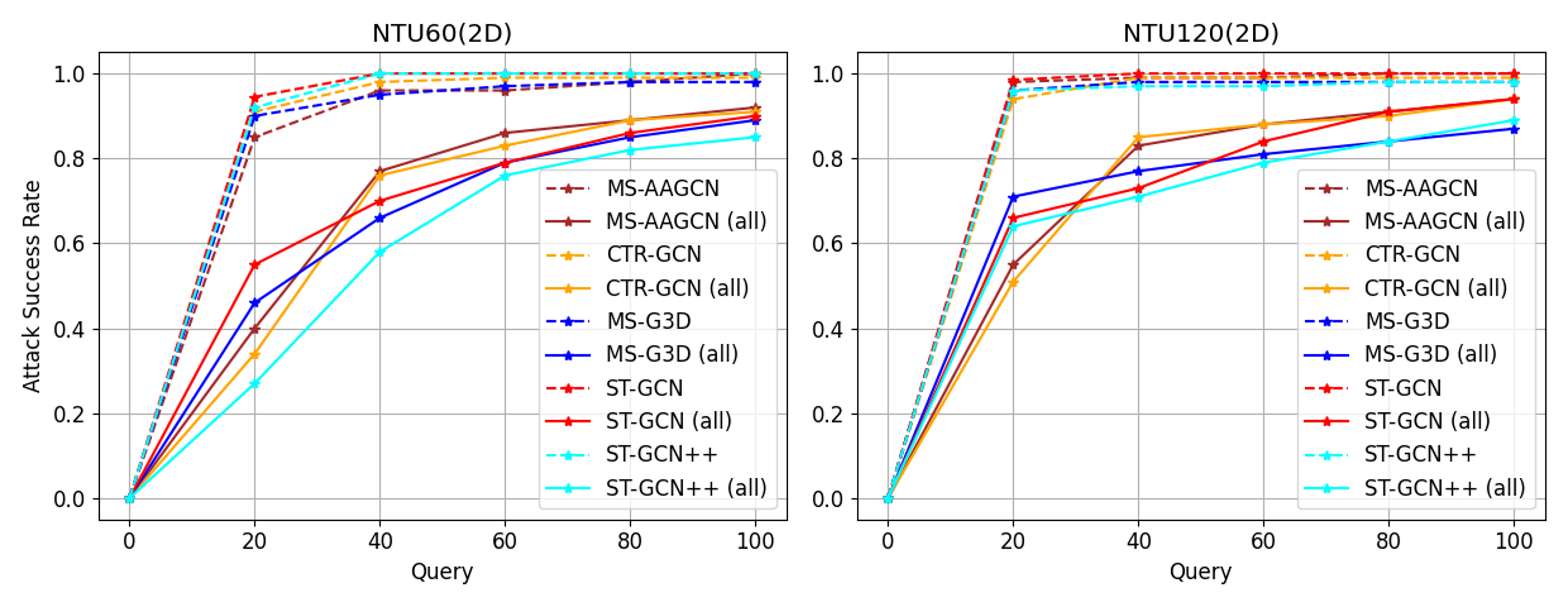}
\end{center}
\vspace{-4mm}
  \caption{Attack performance of ISAAC-K and ISAAC-K (all).}
\label{fig:comparison_with_all}
\vspace{-2mm}
\end{figure}

\noindent\textbf{Key Joint Analyses.}
In order to verify that the key joints extracted from Grad-CAM do positively contribute to the attack, we conduct attacks when key joints are selected randomly. Fig.~\ref{fig:key_joint} shows that, although random selection may be effective, the ASR-query curves converge faster with high ASR as the query increases when the key joints are selected by Grad-CAM. 
In another aspect, Fig.~\ref{fig:comparison_with_all} compares the performance of ISAAC-K with a variant, ISAAC-K (all), where all joints are perturbed. It is not surprising that ISAAC-K performs better than ISAAC-K (all), illustrating that the key joints extracted from Grad-CAM are indeed those that play crucial roles in classification, thus positively contributing to the attack. The gap between ISAAC-K and ISAAC-K (all) also illustrates that sparse perturbations work better.

\subsection{User Study}\label{subsec:user_study}
To measure the indistinguishability of adversarial motions, we conducted a user study on Amazon Turk (AMT)~\cite{amazon_turk}. We recruited 80 human subjects to complete three tasks related to adversarial motions: consistency, naturalness, and realness. In our user study, all questions were optional and we did not collect unnecessary personal information. All participants were over 18 years old and consented for their answers to be used for academic research. Our criterion for selecting participants is that their answer approval rate reaches at least 95\%, indicating that their previous answers were widely recognized by previous survey initiators.
We filter out those arbitrary answers whose answering time is less than eight seconds, since all motions involved in the user study last at least that long. 
The Human Research Ethics Committee of the authors' affiliation determined that the study was exempt from further human subjects review, and we followed best practices for ethical human subjects survey research. 

\noindent\textbf{Survey Protocol.}
1) \textit{Consistency}. Whether the skeletal motion is consistent to the output label plays a dominant role. We randomly selected 10 clean motions and 60 adversarial motions averagely from six GCN-based victim models (10 from each) and provided their original labels and adversarial labels. Subjects were asked to select the label that best describes the skeletal motion. 
2) \textit{Naturalness}. Although we set a restriction threshold on the adversarial motion, it remains difficult to quantitatively assess whether the perturbations are stealthy enough to preserve visual naturalness. Instead, following prior work~\cite{cao2023stylefool}, the subjects were asked to use a Likert-scale~\cite{likert1932technique} from 1 to 5 (higher means more natural) to evaluate the naturalness of randomly selected 60 skeletal motions which consists of 30 clean ones, and 30 adversarial ones. 
3) \textit{Realness}. It is also a challenging task to distinguish which skeletal motion is real (clean) if two motions are played together. The subjects were provided with two motions (a clean one and an adversarial one) with minor differences and asked to choose which skeletal motion is the real one. 

\begin{figure}[t]
\begin{center}
  \includegraphics[width=0.95\linewidth]{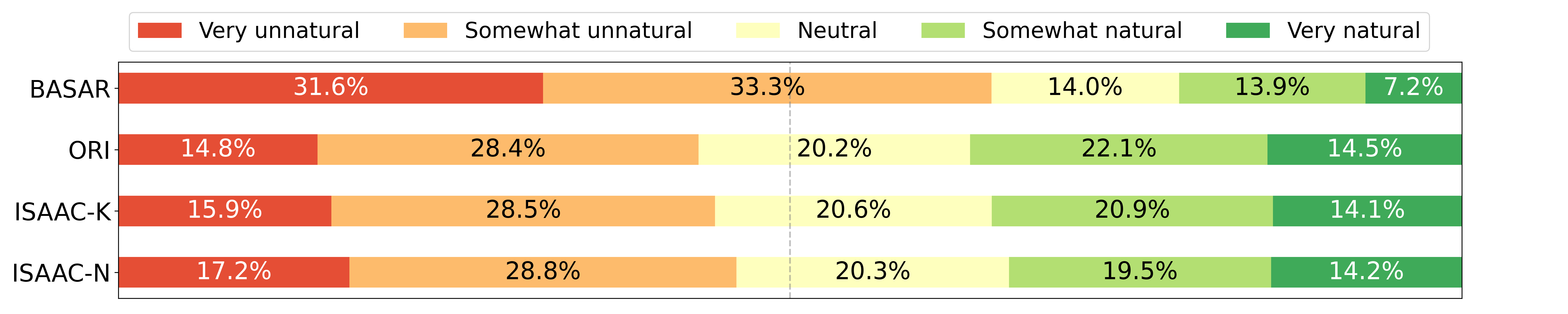}
\end{center}
\vspace{-3mm}
  \caption{Naturalness test results of different attacks.}
\label{fig:naturalness}
\vspace{-1mm}
\end{figure}

\noindent\textbf{Results and Analyses.}
In the consistency test, we find almost all the subjects (92.5\%) chose the original label for adversarial motions, indicating that the perturbations are small enough to preserve the semantic information of the motions.
Fig.~\ref{fig:naturalness} shows the naturalness test results. The subjects gave similar results between clean and adversarial motions. We conduct a Mann-Whitney U-test~\cite{mann1947on} between the two distributions, with the null hypothesis being that there is no significant difference between original and adversarial motions in terms of naturalness. The result (p = 0.08$>$0.05) indicates that ISAAC-K maintains sensory comfort well. Since we modify BASAR to apply to more practical scenarios, the naturalness drops a lot. 
In the realness test, we find that, overall, 32.19\% of the subjects could correctly distinguish adversarial motions from clean ones, which is lower than random guessing. It can be concluded that ISAAC-K can not only maintain semantic information similar to the original motions, but also confuse human subjects visually.

\subsection{Ablation Study}\label{subsec:ablation}
\begin{figure}[t]
\begin{center}
  \includegraphics[width=0.9\linewidth]{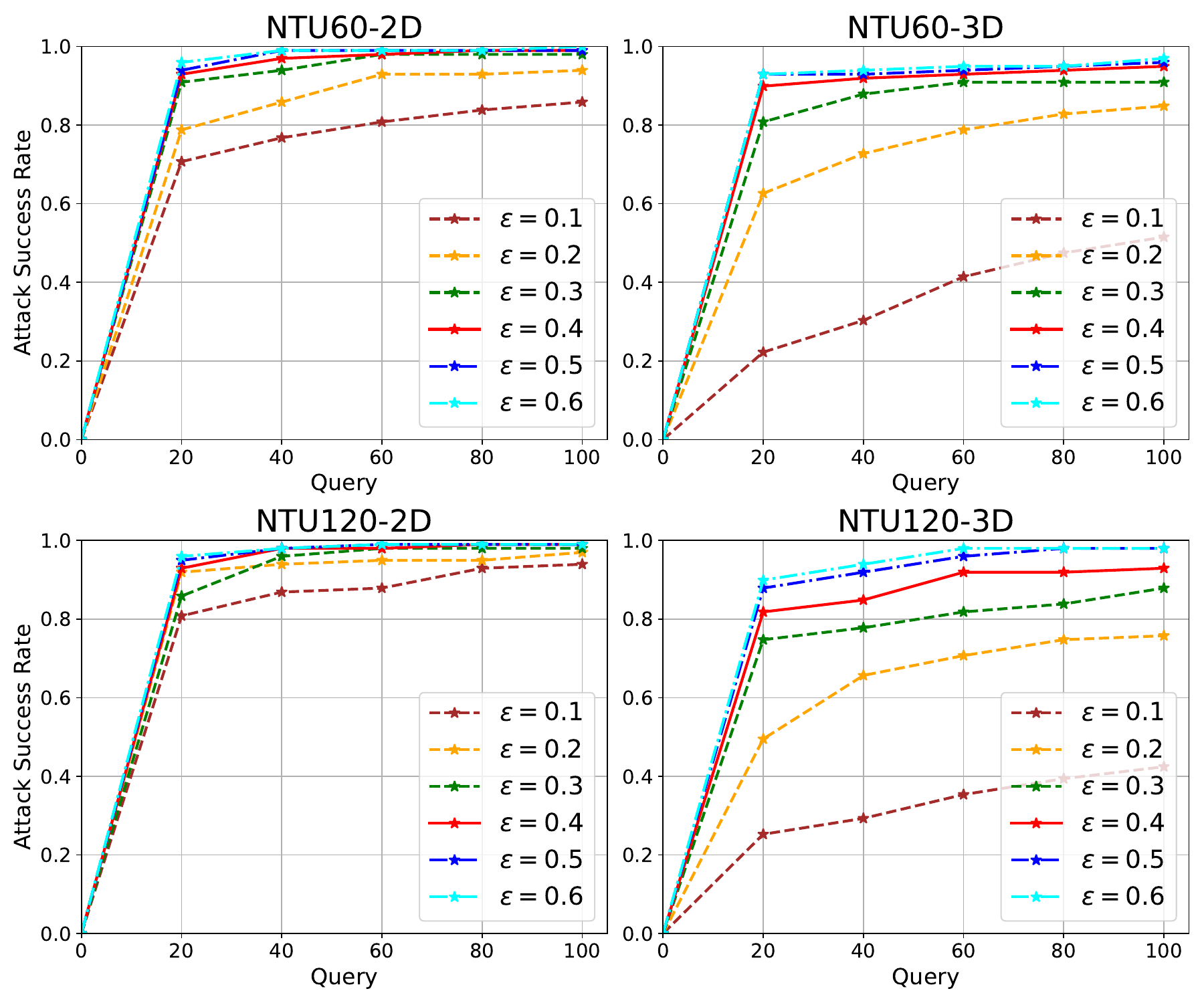}
\end{center}
\vspace{-3mm}
\caption{Attack performance under different restriction thresholds $\varepsilon$.}
\label{fig:ablation_epsilon}
\end{figure}

\begin{figure}[t]
\begin{center}
  \includegraphics[width=0.8\linewidth]{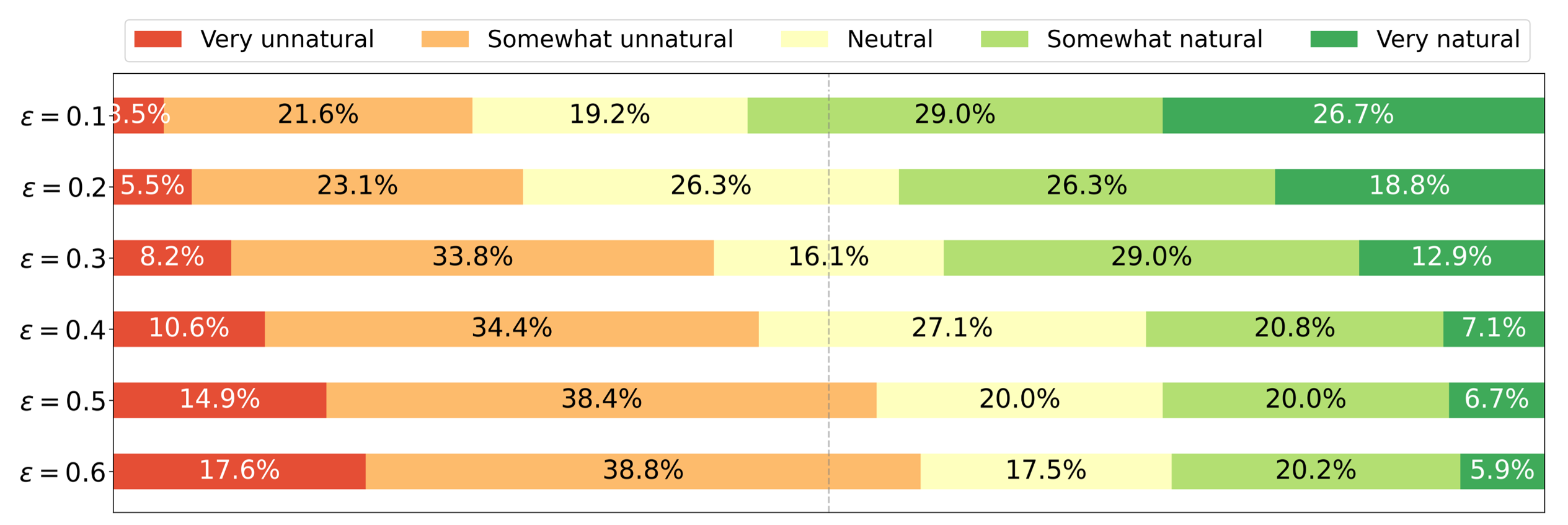}
\end{center}
\vspace{-3mm}
\caption{Naturalness test results under different restriction thresholds $\varepsilon$.}
\label{fig:naturalness_epsilon}
\end{figure}

\noindent\textbf{Restriction Threshold $\varepsilon$.}
Changing the restriction threshold $\varepsilon$ will affect the attack performance as well as, more importantly, the naturalness of the adversarial motions. Fig.~\ref{fig:ablation_epsilon} shows the ASR-query curves under different restriction thresholds $\varepsilon$. ASR increases sharply and then slows down as $\varepsilon$ increases. After $\varepsilon$ exceeds 0.4, ASR increases limitedly and the artifacts become obvious in adversarial motions (the average naturalness drops below ``neutral'' in Fig.~\ref{fig:naturalness_epsilon}), providing a reasonable explanation of why we choose $\varepsilon$ as 0.4 finally.

\noindent\textbf{Key Joint Number $N_k$.}
We also conduct an ablation study under different key joint numbers $N_k$. Fig.~\ref{fig:key_joint} shows the ASR-query curves. The results are consistent with our intuition that a small key joint number will increase the attack difficulty (lower ASR), and a larger key joint number will reduce the attack efficiency (higher AQ). Users can choose different $N_k$ according to their preferences or requirements, given insignificant differences.

\subsection{Attack Performance of ISAAC-N}\label{sec:experiment_ISAAC-N}

We conduct the experiments on a randomly selected subset (500 samples) of the NTU120(3D) dataset. 
The results in Table~\ref{tab:attack_performance_non-semantic} show that ISAAC-NRA performs the best (almost 100\% ASR) due to a larger region of perturbations. ISAAC-NR can also achieve an average of over 20\% ASR, indicating that well-trained models are not robust to simple part-wise replacement. Due to the experimental setup for ISAAC-NR where the attacker has no prior knowledge of the model and cannot query the model, ISAAC-NR can be regarded as a query-free no-box attack.

\begin{figure}[t]
\begin{center}
  \includegraphics[width=1.0\linewidth]{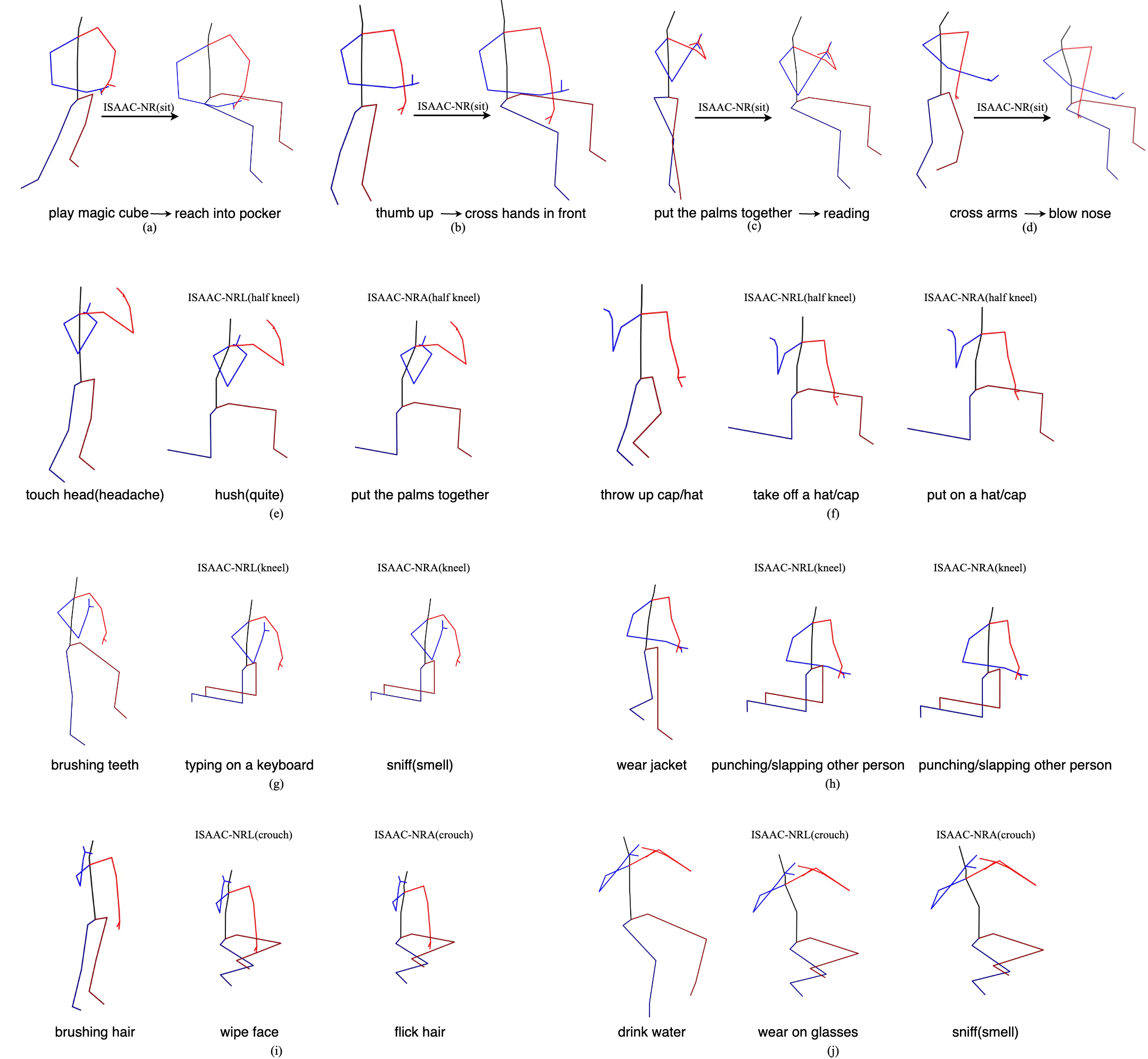}
\end{center}
\vspace{-3mm}
\caption{Some examples of ISAAC-N. The skeleton action recognition model predicts a wrong output towards part-wise replacement.}
\label{fig:example_isaac_n}
\end{figure}

Fig.~\ref{fig:vis}(b) visualizes ISAAC-NR by an actual example. The replacement preserves the semantic information of the motion and maintains the naturalness of the adversarial motion. 
The naturalness test result in Fig.~\ref{fig:naturalness} shows that ISAAC-N obtains sound insidiousness (with a p-value of 0.06$>$0.05) since the perturbations are generated by solely replacing the lower body. Fig.~\ref{fig:example_isaac_n} provides more visualizations for ISAAC-N. 

\begin{table*}[t]  
    \centering
    \caption{Attack performance of ISAAC-N on NTU120(3D).}
    \label{tab:attack_performance_non-semantic}
    \vspace{-1mm}
    \resizebox{0.98\linewidth}{!}{
    \normalsize
    \begin{tabular}{crrrrrrrrrrrrrrrrrrrrrrrrrrrrrrrrrrrrrrr}
    \toprule
    \multirow{3}{*}[-0.5ex]{\textbf{Model}} 
    & \multicolumn{8}{c}{\textbf{ISAAC-NR}}
    & \multicolumn{8}{c}{\textbf{ISAAC-NRL}} 
    & \multicolumn{8}{c}{\textbf{ISAAC-NRA}}
    \\
    \cmidrule(r){2-9}\cmidrule(r){10-17}\cmidrule(r){18-25}
    
    & \multicolumn{2}{c}{\footnotesize{sit}} 
    & \multicolumn{2}{c}{\footnotesize{crouch}} 
    & \multicolumn{2}{c}{\footnotesize{kneel}} 
    & \multicolumn{2}{c}{\footnotesize{half kneel}} 
    & \multicolumn{2}{c}{\footnotesize{sit}} 
    & \multicolumn{2}{c}{\footnotesize{crouch}} 
    & \multicolumn{2}{c}{\footnotesize{kneel}} 
    & \multicolumn{2}{c}{\footnotesize{half kneel}} 
    & \multicolumn{2}{c}{\footnotesize{sit}} 
    & \multicolumn{2}{c}{\footnotesize{crouch}} 
    & \multicolumn{2}{c}{\footnotesize{kneel}} 
    & \multicolumn{2}{c}{\footnotesize{half kneel}} \\
    
    \cmidrule(r){2-3}\cmidrule(r){4-5}\cmidrule(r){6-7}\cmidrule(r){8-9}
    \cmidrule(r){10-11}\cmidrule(r){12-13}\cmidrule(r){14-15}\cmidrule(r){16-17}
    \cmidrule(r){18-19}\cmidrule(r){20-21}\cmidrule(r){22-23}\cmidrule(r){24-25}
    
    & \footnotesize{\textbf{ASR}$\uparrow$} & \footnotesize{\textbf{AQ}$\downarrow$} 
    & \footnotesize{\textbf{ASR}$\uparrow$} & \footnotesize{\textbf{AQ}$\downarrow$} 
    & \footnotesize{\textbf{ASR}$\uparrow$} & \footnotesize{\textbf{AQ}$\downarrow$} 
    & \footnotesize{\textbf{ASR}$\uparrow$} & \footnotesize{\textbf{AQ}$\downarrow$} 
    & \footnotesize{\textbf{ASR}$\uparrow$} & \footnotesize{\textbf{AQ}$\downarrow$} 
    & \footnotesize{\textbf{ASR}$\uparrow$} & \footnotesize{\textbf{AQ}$\downarrow$}
    & \footnotesize{\textbf{ASR}$\uparrow$} & \footnotesize{\textbf{AQ}$\downarrow$} 
    & \footnotesize{\textbf{ASR}$\uparrow$} & \footnotesize{\textbf{AQ}$\downarrow$} 
    & \footnotesize{\textbf{ASR}$\uparrow$} & \footnotesize{\textbf{AQ}$\downarrow$} 
    & \footnotesize{\textbf{ASR}$\uparrow$} & \footnotesize{\textbf{AQ}$\downarrow$} 
    & \footnotesize{\textbf{ASR}$\uparrow$} & \footnotesize{\textbf{AQ}$\downarrow$} 
    & \footnotesize{\textbf{ASR}$\uparrow$} & \footnotesize{\textbf{AQ}$\downarrow$} \\
    \midrule
    MS-AAGCN  & 0.17 & 0.00 & 0.18 & 0.00 & 0.17 & 0.00 & 0.17 & 0.00 & 0.79 & 115.38 & 0.79 & 112.76 & 0.78 & 113.63 & 0.77 & 117.61  & 0.99 & 44.83  & 0.99 & 44.68 & 0.99 & 44.07 & 0.99 & 46.22 \\
    \midrule
    CTR-GCN  & 0.24 & 0.00 & 0.25 & 0.00 & 0.24 & 0.00 & 0.28 & 0.00 & 0.92 & 121.58 & 0.90 & 113.61 & 0.89 & 114.67 & 0.90 & 112.17  & 0.97 & 35.90  & 0.97 & 34.91 & 0.97 & 35.11 & 0.97 & 35.53 \\
    \midrule
    DG-STGCN  & 0.23 & 0.00 & 0.19 & 0.00 & 0.20 & 0.00 & 0.21 & 0.00 & 0.86 & 141.60 & 0.86 & 141.50 & 0.86 & 148.53 & 0.86 & 150.54  & 0.99 & 58.49  & 0.99 & 56.04 & 0.99 & 57.52 & 0.99 & 59.17  \\
    \midrule
    MS-G3D  & 0.37 & 0.00 & 0.36 & 0.00 & 0.33 & 0.00 & 0.34 & 0.00 & 0.93 & 113.72 & 0.93 & 111.92 & 0.94 & 133.66 & 0.94 & 118.11  & 0.99 & 39.37  & 0.99 & 43.12 & 0.99 & 39.71 & 0.99 & 42.86  \\
    \midrule
    ST-GCN  & 0.05 & 0.00 & 0.06 & 0.00 & 0.05 & 0.00 & 0.07 & 0.00 & 0.36 & 190.26 & 0.35 & 185.32 & 0.35 & 188.27 & 0.34 & 163.19  & 0.99 & 58.48  & 0.99 & 56.44 & 0.99 & 58.18 & 0.99 & 58.72  \\
    \midrule
    ST-GCN++  & 0.18 & 0.00 & 0.19 & 0.00 & 0.22 & 0.00 & 0.21 & 0.00 & 0.76 & 200.90 & 0.75 & 185.11 & 0.78 & 217.99 & 0.75 & 204.53  & 0.98 & 63.15  & 0.98 & 71.23 & 0.98 & 66.07  & 0.98 & 73.38 \\
    \bottomrule
    \end{tabular}
    }
    \vspace{-1mm}
\end{table*}

\subsection{Countermeasure}\label{subsec:countermeasure}
\noindent\textbf{Performance against the Existing Defense.}
To test the ability of our attack to resist defense, we employ a SOTA skeletal defense, BEAT~\cite{wang2023defending}, which is a Bayesian-based AT method. 
We do not validate our attack on 2D data, since BEAT is tailored for 3D data. Table~\ref{tab:countermeasure} reports the performance of re-attacking the trained defense models by ISAAC-K. Due to the differences in perturbation constraints and sparsity levels between the adversarial examples used in BEAT and those generated by ISAAC-K, the ASR and AQ are close to those before adversarial training, indicating that BEAT is almost unable to defend against ISAAC-K. Considering the performance of BASAR under BEAT (verified in the paper of BEAT~\cite{wang2023defending}), we find that BEAT can alleviate BASAR, but BEAT's ability to defend against ISAAC-K is very limited. One possible reason could be that BASAR and BEAT are a combination of attack and defense that focuses on perturbation stealthiness rather than attack efficiency.

\begin{table}[t]
    \caption{Performance of ISAAC-K against BEAT and RS.}
    \label{tab:countermeasure}
    \centering
    \vspace{-1mm}
    \resizebox{0.78\linewidth}{!}{
    \begin{tabular}{ccrr|c|rr|c}
    \toprule
    \multirow{2}{*}[-0.5ex]{\textbf{Model}} & \multirow{2}{*}[-0.5ex]{\textbf{Defense}} & \multicolumn{3}{c}{\textbf{NTU60(3D)}} & \multicolumn{3}{c}{\textbf{NTU120(3D)}} \\
    \cmidrule(r){3-5}\cmidrule(r){6-8}
    & & \textbf{ASR}$\uparrow$ & \textbf{AQ}$\downarrow$ & \textbf{DSR}$\uparrow$ & \textbf{ASR}$\uparrow$ & \textbf{AQ}$\downarrow$ & \textbf{DSR}$\uparrow$ \\
    \midrule
    \multirow{2}{*}{MS-AAGCN} & Before & 0.99 & 29.55 & \multirow{2}{*}{0.01} & 0.97 & 30.73 & \multirow{2}{*}{0.00} \\
    & After &0.99 &37.20 & &0.97 &32.15 & \\
    \midrule
    \multirow{2}{*}{CTR-GCN} & Before & 0.99 & 27.44 & \multirow{2}{*}{0.01} & 0.98 & 34.11 & \multirow{2}{*}{0.03} \\
    & After &0.99 &29.05 & &0.97 &34.35 & \\
    \midrule
    \multirow{2}{*}{DG-STGCN} & Before & 0.99 & 38.31 & \multirow{2}{*}{0.01} & 1.00 & 43.42 & \multirow{2}{*}{0.07} \\
    & After &0.99 &35.04 & &1.00 &90.65 & \\
    \midrule
    \multirow{2}{*}{MS-G3D} & Before & 0.99 & 27.25 & \multirow{2}{*}{0.02} & 0.98 & 35.29 & \multirow{2}{*}{0.04} \\
    & After &0.98 &30.23 & &0.98 &37.33 & \\
    \midrule
    \multirow{2}{*}{ST-GCN} & Before & 1.00 & 20.26 & \multirow{2}{*}{0.00} & 1.00 & 22.60 & \multirow{2}{*}{0.01}\\
    & After  &0.99 &23.05 &  &1.00 &25.28 & \\
    \midrule
    \multirow{2}{*}{ST-GCN++} & Before & 1.00 & 34.71 & \multirow{2}{*}{0.01} & 0.98 & 30.26 & \multirow{2}{*}{0.02}\\
    & After &0.99 &30.03 & &0.98 &41.74 & \\
    \bottomrule
    \end{tabular}
    }
    \vspace{-1mm}
\end{table}

Furthermore, we evaluate the robustness of our attacks against a defense based on randomized smoothing (RS)~\cite{zheng2020towards}. We extend the RS framework from image classification to skeleton action recognition by incorporating Gaussian noise and temporal filtering. This method computes a certified robustness radius based on the variance of perturbed samples, and considers an adversarial skeleton to be defended if its distance from the clean skeleton lies within this radius. As shown in Table~\ref{tab:countermeasure}, the Defense Success Rate (DSR) of RS is consistently close to 0 against our attacks. This can be attributed to two factors: (1) our perturbations are crafted under an $\ell_\infty$-norm constraint, for which RS does not provide certification guarantees~\cite{kumar2020curse}; and (2) skeleton data is inherently sparse, topologically discrete, and low-dimensional. Therefore, skeleton models are less sensitive to small perturbations than image models. 
These characteristics limit the effectiveness of RS-based certification on skeleton models.

\noindent\textbf{Adaptive Defense.}
We also show that our proposed attack can in return help improve the adversarial robustness. 
As AT has been universally acknowledged as the most feasible defense in theory, we ameliorate AT and propose several adaptive defenses based on adversarial motions generated by ISAAC-K and ISAAC-N. 
Generally, given dataset $\mathcal{D}$ and cross-entropy loss function $\widetilde{\ell}$, we solve the following min-max problem for model $f$ with parameters $\theta$.
\begin{equation}
\small
\mathop {\min }\limits_\theta  {\mathop \mathbbm{E}\limits_{\left( {{\mathbf{x}},y} \right)\sim \mathcal{D}}}\left[ {\mathop {\max }\limits_{r,{\mathbf{d}}} \widetilde{\ell}\left[ {\Theta \left( {{\varphi \left( {\mathbf{x}} \right)} + r\delta\frac{{{M_x} \odot {\mathbf{d}}}}{{{{\left\| {{M_x} \odot {\mathbf{d}}} \right\|}_2}}},{\varphi \left( {\mathbf{x}} \right)},{\varepsilon _b}} \right),y} \right]} \right].
\end{equation}

Let $\mathcal{R}$ denote the replacement operation given replacement mask matrix $M_r$ and pre-designed motion $\mathbf{x}_r$. We provide four versions of ISAAC-based AT strategies:

\begin{enumerate}
\item  AT-ISAAC-K: $\varphi \left( {\mathbf{x}} \right) = {\mathbf{x}},\delta  = 1,{M_x} = {M}$.

\item  AT-ISAAC-NR: $\varphi \left( {\mathbf{x}} \right) = {\mathcal{R}}\left( {{\mathbf{x}},{{\mathbf{x}}_r},{M_r}} \right)$, $\delta  = 0$, $\mathop {\max }\limits_{r,{\mathbf{d}}}  \to \mathop {\max }\limits_{{{\mathbf{x}}_r},{M_r}} $.

\item  AT-ISAAC-NRL: $\varphi \left( {\mathbf{x}} \right) = {\mathcal{R}}\left( {{\mathbf{x}},{{\mathbf{x}}_r},{M_r}} \right)$, $\delta  = 1$, ${M_x} = {M_r}, \mathop {\max }\limits_{r,{\mathbf{d}}}  \to {\mathbbm{E}_{\left( {{{\mathbf{x}}_r},{M_r}} \right)}}\mathop {\max }\limits_{r,{\mathbf{d}}} $.

\item  AT-ISAAC-NRA: $\varphi \left( {\mathbf{x}} \right) = {\mathcal{R}}\left( {{\mathbf{x}},{{\mathbf{x}}_r},{M_r}} \right)$, $\delta  = 1$, ${M_x} = \mathbf{1}_{T \times N \times D}, \mathop {\max }\limits_{r,{\mathbf{d}}}  \to {\mathbbm{E}_{\left( {{{\mathbf{x}}_r},{M_r}} \right)}}\mathop {\max }\limits_{r,{\mathbf{d}}} $.
\end{enumerate}

To characterize how our designed defenses influence model generalization under structured adversarial perturbations, we derive a unified robust generalization bound based on the Rademacher complexity~\cite{yin2019rademacher}. 
Given a hypothesis class $\mathcal{F}$, for any $\eta \in (0,1)$, with probability at least $1-\eta$, the following inequality holds for all $f\in\mathcal{F}$,
\begin{equation}
\small
\widetilde{R}(f) \leq \widetilde{R}_n(f) + 2\mathfrak{R}_{\mathcal{S}}(\widetilde{\ell}_{\mathcal{F}}) + 3 \sqrt{ \frac{ \log(2/\eta) }{2|\mathcal{D}|} },
\end{equation}
where $\widetilde{R}(f)$ denotes the population risk, and $\widetilde{R}_n(f)$ denotes the empirical risk computed over the training dataset $\mathcal{D}$.
The term $\mathfrak{R}_{\mathcal{S}}(\widetilde{\ell}_{\mathcal{F}})$ denotes the adversarial Rademacher complexity of the loss class $\widetilde{\ell}_{\mathcal{F}}$, and is upper bounded by: $\xi \cdot \varepsilon \cdot \sup_{\mathbf{x}} \| \nabla f(\mathbf{x}) \|_{\infty, k_x} \cdot \sqrt{ \frac{ k_x \cdot \log d }{|\mathcal{D}|} }$, where 
$\xi$ is a constant that accounts for the Lipschitz continuity of the loss function with respect to the model output, 
$k_x = \| M_x \|_0$ is the number of perturbed input dimensions, and
$\| \nabla f(\mathbf{x}) \|_{\infty, k_x}$ denotes the maximum top-\(k_x\) gradient norm reflecting the model's local sensitivity. 
This bound shows that reducing the number of perturbed dimensions $k_x$ can lead to tighter generalization guarantees, thereby revealing a trade-off between adversarial coverage and sample complexity. 

\begin{table}[t]  
    \centering
    \caption{Adaptive defense performance.}
    \label{tab:defense_tab}
    \vspace{-1mm}
    \resizebox{0.65\linewidth}{!}{
    \begin{tabular}{lrrrr}
    \toprule
    \multicolumn{1}{c}{\multirow{2}{*}{\textbf{Method}}} & \multicolumn{2}{c}{Before}                     & \multicolumn{2}{c}{After}                      \\ 
    \cmidrule(r){2-3}\cmidrule(r){4-5}
    \multicolumn{1}{c}{}                         & \textbf{ASR}$\uparrow$ & \textbf{AQ}$\downarrow$ & \textbf{ASR}$\uparrow$ & \textbf{AQ}$\downarrow$ \\ 
    \midrule
    AT-ISAAC-K                                     & \multicolumn{1}{l}{0.98}         & 30.26      & \multicolumn{1}{l}{0.90}         & 38.20      \\ 
    \midrule
    AT-ISAAC-NR                                    & \multicolumn{1}{l}{0.18}             &0.00             & \multicolumn{1}{l}{0.12}             & 0.00             \\ 
    \midrule
    AT-ISAAC-NRL                                   & \multicolumn{1}{l}{0.76}         & 200.90           & \multicolumn{1}{l}{0.65}             & 236.72              \\ 
    \midrule
    AT-ISAAC-NRA                                   & \multicolumn{1}{l}{0.98}             & 63.15            & \multicolumn{1}{l}{0.86}             & 79.01           \\ 
    \bottomrule
    \end{tabular}
    }
    \vspace{-1mm}
\end{table}

We train four ST-GCN++ models with AT, and we only consider the lower body of ``sit'' in the last three versions. During training, we use examples optimized by 5 PGD steps~\cite{madry2018towards}. 
The results in Table~\ref{tab:defense_tab} demonstrate a certain defense capability of these four adaptive defenses.
After AT, ASR drops in all defenses, among which AT-ISAAC-NRA contributes the most to ASR reduction. AQ also increases after AT, indicating the increasing difficulty for re-attacks. However, we find that AT still remains insufficient for our attacks. One possible reason is that, unlike image models where 5 PGD steps usually push samples close to the decision boundary, our attacks on sparse and low-dimensional skeleton data with additional kinematic and consistency constraints require much deeper optimization to make the skeleton truly adversarial. Consequently, the PGD samples used during training often remain far from the decision boundary, and the mismatch between their distribution and that of our attacks further limits the effectiveness of AT.

\subsection{Extension to Targeted Attacks}\label{subsec:targeted_attacks}

\noindent\textbf{Targeted Attacks for ISAAC-K.}
We also conduct targeted attacks for ISAAC-K. According to our experiments, the original search method of RayS works well enough for untargeted attacks, and its advantages in the $\ell_\infty$ setting are well adapted to our requirements for motion naturalness. However, 
RayS is known to perform poorly in targeted attacks since there is no appropriate initial search direction for the original motion. Therefore, we propose a search direction optimization here. Following query-based targeted attacks~\cite{ilyas2018black}, we start the targeted attack from a motion selected in the target class and take the direction from the target motion to the original motion as the initial direction. This direction enables the search to reach the target area faster.
Also, as an improvement of RayS, we do not start the search with a radius of $\infty$. Instead, we replace it with the decision boundary distance of the initial direction. This will reduce the search space to avoid redundant searches in the early stage. Overall, we start the attack from a target class motion $\mathbf{x}_t$, and optimize the search direction by setting $\mathbf{d}_{best} \leftarrow M \odot {\mathop{\rm sgn}} \left( \mathbf{x}_t - \mathbf{x} \right)$ and ${r_{best}} \leftarrow {g_{res}}\left( {{{\mathbf{x}}_t} - {\mathbf{x}}} \right)$.

With regard to the experiments, our experimental setting is different from that of BASAR. In BASAR, the target class motion can be replaced based on the classification results until an appropriate motion is obtained (which results in their 100\% ASR), but this is not practical. Therefore, to ensure fairness, we randomly select a target class motion when implementing BASAR.
We conduct targeted attacks on the NTU60(2D) and NTU120(2D) dataset. 
Table~\ref{tab:attack_performance_targeted} reports the attack performance on five models (we omit DG-STGCN since the model for 2D data is not available). The results show that, in most cases, ISAAC-K performs better than BASAR in all four metrics, indicating that ISAAC-K also performs well in targeted attacks. ISAAC-K bridges a huge gap in terms of ASR and AQ across different datasets and black-box models. In particular, the AQ has been reduced by an order of magnitude when attacking MS-AAGCN, and the ASR is also higher. An interesting dicovery is that the $l_c$ and $\Delta a$ of ISAAC-K in targeted attacks are better than those in untargeted attacks. One possible reason might be that the search direction optimization leads to fewer queries in targeted attacks (indicating smaller perturbation amplitude). Also note that AQ only considers those samples that have been successfully attacked. 

\begin{table}[t]
    \centering
    \caption{Attack performance comparison of targeted ISAAC-K.}
    \label{tab:attack_performance_targeted}
    \resizebox{0.99\linewidth}{!}{
    \begin{tabular}{ccrrrrrrrrrrrrrrrrrrrrrrrrr}
    \toprule
    \multirow{2}{*}[-0.5ex]{\textbf{Model}} & 
    \multirow{2}{*}[-0.5ex]{\textbf{Attack}} & \multicolumn{4}{c}{\textbf{NTU60(2D)-T}} & \multicolumn{4}{c}{\textbf{NTU120(2D)-T}} \\ 
    \cmidrule(r){3-6}\cmidrule(r){7-10}
    & & \footnotesize{\textbf{ASR}} & \footnotesize{\textbf{AQ}} & \footnotesize{\textbf{$l_c$}} & \footnotesize{\textbf{$\Delta_a$}} 
    & \footnotesize{\textbf{ASR}} & \footnotesize{\textbf{AQ}} & \footnotesize{\textbf{$l_c$}} & \footnotesize{\textbf{$\Delta_a$}} \\
    \midrule
    \multirow{2}{*}{MS-AAGCN} & BASAR~\cite{diao2021basar} & 0.09 & 123.02 & 0.12 & 2.17 
    & 0.08 & 168.93 & 0.11 & 2.54 \\
     & ISAAC-K  
    & \textbf{0.21} & \textbf{15.86} & 0.08 & 1.67 
    & \textbf{0.26} & \textbf{14.73} & 0.06 & 1.41 \\
    \midrule
    \multirow{2}{*}{CTR-GCN} & BASAR~\cite{diao2021basar} & 0.09 & 122.8 & 0.10 & 3.02 
    & 0.17 & 102.97 & 0.08 & 2.44 \\
    & ISAAC-K  
    & \textbf{0.23} & \textbf{21.96} & 0.06 & 1.67 
    & \textbf{0.26} & \textbf{16.46} & 0.09 & 1.46 \\
    \midrule
    \multirow{2}{*}{MS-G3D} & BASAR~\cite{diao2021basar} & 0.06 & \textbf{6.20} & 0.15 & 1.96 
    & 0.09 & \textbf{11.72} & 0.09 & 2.20 \\
    & ISAAC-K  
    & \textbf{0.19} & 14.74 & 0.07 & 1.61 
    & \textbf{0.26} & 14.62 & 0.05 & 1.35 \\
    \midrule
    \multirow{2}{*}{ST-GCN} & BASAR~\cite{diao2021basar} & 0.06 & 105.67 & 0.09 & 3.26 
    & 0.11 & 159.11 & 0.09 & 2.43 \\
    & ISAAC-K  
    & \textbf{0.24} & \textbf{15.12} & 0.06 & 1.26 
    & \textbf{0.31} & \textbf{14.65} & 0.07 & 1.63 \\
    \midrule
    \multirow{2}{*}{ST-GCN++} & BASAR~\cite{diao2021basar} & \textbf{0.30} & 368.63 & 0.08 & 3.15 
    & 0.05 & \textbf{6.70} & 0.15 & 1.86 \\
    & ISAAC-K  
    & 0.22 & \textbf{14.73} & 0.04 & 1.47 
    & \textbf{0.23} & 14.70 & 0.07 & 1.47 \\
    \bottomrule
    \end{tabular}
    }
\end{table}

\noindent\textbf{Targeted Attacks for ISAAC-N.}
We only conduct untargeted attacks for ISAAC-N in the main experiment since devising a comprehensive attack from the perspective of non-semantic regions is not the focus of this paper. Nonetheless, we provide a somewhat feasible idea to conduct targeted attacks.
Specifically, we can simply divide the human body into six regions according to joints, \ie the left hand, the right hand, the left leg, the right leg, the spine and the head. According to the number of $N_n \left( = N - N_k \right)$ joints belonging to the above six regions, we can obtain one or two non-semantic regions and replace them as a whole, \ie replacing the sitting posture in both leg areas with kneeling. 
For targeted attacks, one possible solution is to first collect sets of key joints in different regions from the target class, and then select a motion $\mathbf{x}_t$ from the set according to the region to be replaced by the sample to be attacked and then replace it in the following way: 
\begin{equation}\label{equ:non-semantic attack}
\small
{\mathbf{x}_{rep}} = \Psi \left( {\mathbf{x} \odot \left( {1 - {M_r}} \right),{\mathbf{x}_t} \odot {M_r}} \right),
\end{equation}
where ${\mathbf{x}_{rep}}$ denotes the replaced motion, and $\Psi$ denotes the splicing operation of two motions, including the translation of the root joint and the normalization of the motion. Then, minor perturbations can be added until the motion is misclassified to the target class. We conduct targeted attacks on NTU120(3D) and choose ``sit'' as the target. Results in Table~\ref{tab:attack_performance_isaac-n_targeted} indicate that ISAAC-N can achieve certain targeted performance, thereby enhancing its applicability to more realistic threat scenarios.

\begin{table}[t]  
    \centering
    \caption{Attack performance for targeted ISAAC-N.}
    \label{tab:attack_performance_isaac-n_targeted}
    \vspace{-1mm}
    \resizebox{0.8\linewidth}{!}{
    \normalsize
    \begin{tabular}{crrrrrrrrrrrrrrrrrr}
    \toprule
    \multirow{2}{*}[-0.5ex]{\textbf{Model}} 
    & \multicolumn{2}{c}{\textbf{ISAAC-NR}}
    & \multicolumn{2}{c}{\textbf{ISAAC-NRL}} 
    & \multicolumn{2}{c}{\textbf{ISAAC-NRA}}
    \\
    \cmidrule(r){2-3}\cmidrule(r){4-5}\cmidrule(r){6-7}
    & \footnotesize{\textbf{ASR}$\uparrow$} & \footnotesize{\textbf{AQ}$\downarrow$} 
    & \footnotesize{\textbf{ASR}$\uparrow$} & \footnotesize{\textbf{AQ}$\downarrow$} 
    & \footnotesize{\textbf{ASR}$\uparrow$} & \footnotesize{\textbf{AQ}$\downarrow$} \\
    \midrule
    MS-AAGCN  & 0.05 & 0.00 & 0.08 & 6.38 & 0.49 & 15.10  \\
    \midrule
    CTR-GCN  & 0.05 & 0.00 & 0.09 & 7.33 & 0.38 & 15.55 \\
    \midrule
    DG-STGCN  & 0.00 & 0.00 & 0.00 & 0.00 & 0.07 & 15.71 \\
    \midrule
    MS-G3D  & 0.08 & 0.00 & 0.13 & 6.38 & 0.45 & 14.82\\
    \midrule
    ST-GCN  & 0.14 & 0.00 & 0.19 & 4.84  & 0.37 & 14.81\\
    \midrule
    ST-GCN++ & 0.13 & 0.00 & 0.17 & 4.41 & 0.45 & 14.89 \\
    \bottomrule
    \end{tabular}
    }
    \vspace{-1mm}
\end{table}

\subsection{Surrogate Model Analyses}\label{subsec:surrogate}
\begin{figure}[t]
\begin{center}
  \includegraphics[width=0.85\linewidth]{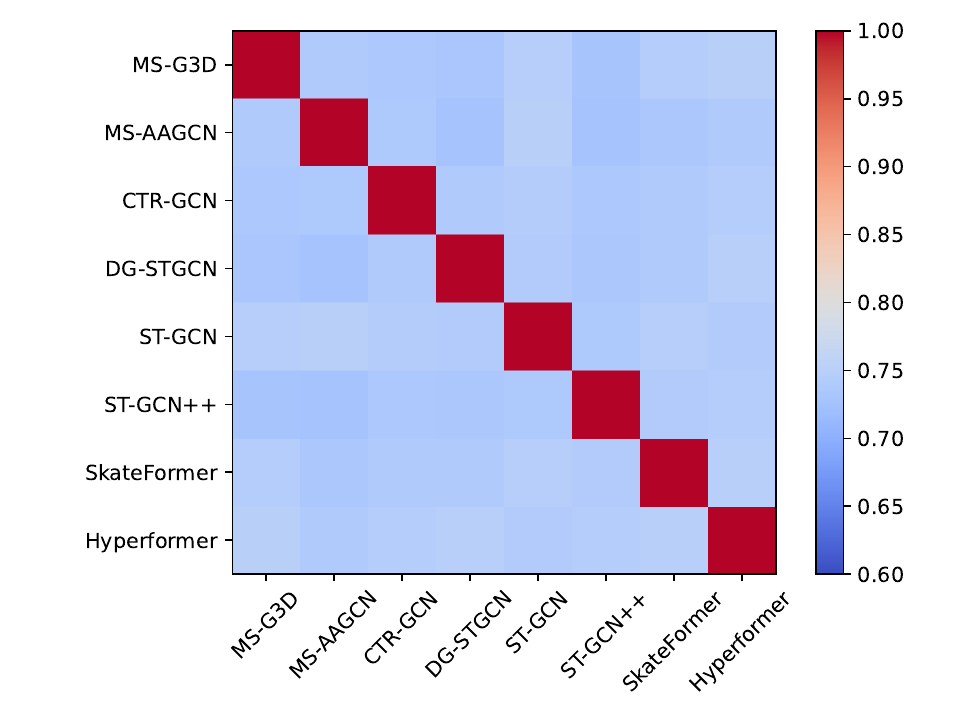}
\end{center}
\vspace{-3mm}
\caption{Cosine similarity for key joints extracted from different surrogate models.}
\label{fig:sim}
\end{figure}

\noindent\textbf{Dependency on Surrogate Model.}
In our pipeline, the surrogate model is used only to extract key joints that guide the placement of sparse perturbations. This step does not require the surrogate to resemble the victim model, since it merely approximates motion-sensitive regions. In realistic black-box scenarios, attackers can freely choose any open-source skeletal model for this purpose, since only the victim model is inaccessible. While Grad-CAM may yield varying key joints across different surrogate models, this variability is unlikely to substantially affect attack performance. This conclusion is supported by a pilot experiment that assesses the cosine similarity of the obtained key joints for eight models on NTU60(3D). As shown in Fig.~\ref{fig:sim}, the key joints are similar among models, indicating that changing the surrogate model has little impact on the output results. It is observed that ST-GCN achieves the highest overall key joint similarity with other models, which also provides support for choosing ST-GCN as the surrogate model in our experiments.

\noindent\textbf{Role of the Surrogate Model.}
The role of the surrogate in our method differs fundamentally from that in traditional  gray-box (transfer-based) attacks. Prior work typically optimizes perturbations in a white-box manner on a surrogate and then transfers them to the victim model, which makes the attack highly dependent on the surrogate architecture and often results in low transferability, \eg perturbations optimized for the surrogate may fall back inside the decision boundary of the victim model and fail to cause misclassification. In contrast, our method treats the surrogate model solely as a tool for extracting key joints used to guide the perturbation process. The actual attack optimization is carried out entirely in the black-box setting by directly querying the victim model. This results in a pure black-box attack that does not rely on architectural similarity between the surrogate and the victim model, and is thus significantly more robust to model mismatch.

\section{Discussion}\label{sec:discussion}
\noindent\textbf{Input Feature Dimension.}
In skeleton-based action recognition, most mainstream models adopt a consistent input format, typically 25 joints, due to the standardization established by the NTU 60/120 datasets. Captured using Kinect V2, which detects 25 body joints, these datasets have become the dominant benchmark, and most models, including GCN-based and ViT-based architectures, are designed accordingly. Therefore, the issue of dimensional mismatch between surrogate and victim models rarely arises. That said, if, in the future, new sensing hardware springs out and leads to dominant datasets with fine-grained skeletal structures (\eg 30 joints), we believe that new standardized formats are likely to emerge, with models adjusted to support the updated input. Even in the absence of a unified format, a surrogate model can be adapted to match the victim model's input. For example, one could start with a well-performing 25-joint model, adjust its early layers to accept 30-joint data, and train it to mimic the outputs of the victim model. In this way, the surrogate model remains aligned in structure and output behavior, allowing key joint extraction to proceed as in our proposed attack. 

\noindent\textbf{Long Videos as Input.}
We use the count-based key joint selection since videos in our dataset mostly contain only one action. If the input video is long and contains more actions, the key joints of different frames may appear different. 
As a possible alternative solution, the motion $\mathbf{x}$ can be split into $K$ sub-action segments $\left\{ {{S_1},{S_2},...,{S_K}} \right\}$, where ${S_i} = \left\{ {{\mathbf{x}_{{t_i}}},{\mathbf{x}_{{t_i} + 1}},...,{\mathbf{x}_{{t_i} + {T_i} - 1}}} \right\}$, $\mathbf{x}_{t_i}$ represents the ${t_i}$-th frame of the motion, ${t_i}$ and ${T_i}$ denote the first frame index and the frame number of the $i$-th segment, respectively. Then the key joints are extracted respectively among different segments but remain the same in each segment.

\noindent\textbf{Other Scenarios of Inputs.}
For video inputs, there exist mature methods~\cite{zheng2023deep} for extracting keypoints from videos. In existing skeleton datasets, the 3D coordinates are estimated from the 2D data. Although keypoint extraction is becoming increasingly refined, this does not mean that more keypoints lead to more effective action recognition; rather, it poses greater challenges~\cite{song2021human}, and can serve as a potential direction for future research.

\noindent\textbf{Social Impact.}
As an adversarial attack, our proposed method may be maliciously or unintentionally used to threaten the security of skeleton recognition systems in the real world. 
We hope that our work can arouse the related community to attach greater significance to the vulnerability of skeletal models, since their accuracy, as well as robustness, are crucial to widely used scenarios, such as aged care, video surveillance, \etc. We believe that our discovery of a more efficient and effective sparse attack, as well as the non-semantic attack, will be accorded due significance and treated as a novel respect to help improve the robustness of skeletal models in steeply rising safety-critical systems. 

\section{Conclusion}
This paper chalks out undermining reliance on skeleton recognition systems with higher efficiency. We orchestrate an efficient skeletal adversarial attack called ISAAC-K under the black-box setting. By considering perturbations only on key joints, we capitalize on Grad-CAM to extract dominant joints from a local surrogate model and then optimize the perturbation under intra- and inter-frame constraints using an improved version of RayS. We show that ISAAC-K can achieve better attack performance over BASAR while maintaining indistinguishability, which is verified by the judgment of human subjects in the user study. Additionally, we unintentionally find that replacing some parts of the skeleton can also affect the model output, which provides us with a new angle - adding large perturbations to non-semantic regions - to launch a query-free no-box attack. Based on these attacks, we also propose four adaptive defenses to help skeleton action recognition models focus more on semantic contents, and thus improve the robustness of these models. Future work includes researching more powerful and generalizable defense methods.

\section*{Acknowledgments}
The authors thank the anonymous reviewers for their valuable feedback that helped improve the paper. This work was supported in part by National University of Singapore (No.A-8003872-00-00), the University of Hong Kong and Ping An Technology. This work was also partially supported by the NSFC for Young Scientists of China (No.62202400) and the RGC for Early Career Scheme (No.27210024). Any opinions, findings, or conclusions expressed in this material are those of the authors and do not necessarily reflect the views of NSFC and RGC.

{\small
\bibliographystyle{IEEEtran}
\bibliography{main}
}

\end{document}